\documentclass[aps,prc,twocolumn,superscriptaddress,longbibliography,floatfix]{revtex4-2}

\usepackage[T1]{fontenc}

\usepackage{amsmath,amssymb,graphicx,booktabs,bm}

\usepackage[colorlinks=true,citecolor=blue,urlcolor=blue,linkcolor=blue]{hyperref}

\graphicspath{{figures/}}

\newcommand{\gef}{\textsc{gef}}

\begin{document}

\title{Probabilistic Machine Learning for Total Kinetic Energy Distributions in Neutron-Induced Fission}

\author{Vaia Prassa}

\email{vprassa@uth.gr}

\affiliation{Department of Physics, University of Thessaly, 35100 Lamia, Greece}

\author{Martha Papadopoulou}

\affiliation{Department of Physics, Aristotle University of Thessaloniki, 54124 Thessaloniki, Greece}

\author{Christos Bompotas}

\affiliation{Department of Physics, Aristotle University of Thessaloniki, 54124 Thessaloniki, Greece}

\date{\today}

\begin{abstract}

\begin{description}

\item[Background] The incident-neutron-energy dependence of post-neutron total-kinetic-energy (TKE) distributions, particularly their width in the multichance-fission regime, provides a sensitive test of fission calculations. Experimental TKE distributions at higher neutron energies remain comparatively sparse.

\item[Purpose] We investigate how the General Description of Fission Observables (GEF) and probabilistic machine-learning representations trained on GEF reproduce measured post-neutron TKE distributions, with particular emphasis on high-energy width predictions and behavior outside the incident-energy range represented in training.

\item[Methods] Gaussian process regression (GPR), Gaussian process regression with a white-noise kernel (GPR-WK), a mixture density network (MDN), and a Bayesian mixture density network (BMDN) are first evaluated on a common dataset of calculated post-neutron TKE distributions. An extended MDN/BMDN analysis uses 1,025 GEF spectra from 20 target nuclei and 60 measured distributions. We compare training data restricted to $E_n\le20$ MeV, all available spectra through 40 MeV, and the full range ($\sim76$ MeV), together with matched-size controls that extend the training range to 30 or 40 MeV while preserving the number of complete spectra for each target.

\item[Results] Extending the calculated training range substantially improves the high-energy TKE-distribution width. For BMDN above 40 MeV, the width mean absolute error (MAE) relative to experiment decreases from 5.63 MeV for training restricted to \(E_n\le20\) MeV to 2.70 and 2.53 MeV for two matched-size \(E_n\le30\) MeV controls and 1.77 MeV for the matched-size \(E_n\le40\) MeV case. Comparison with the 60 measured spectra shows target-dependent agreement in both the mean TKE and distribution width, without a uniform ordering among GEF, MDN, and BMDN across the five nuclei. Above 40 MeV, BMDN predictive-interval coverage relative to GEF rises from 52.3\% for the 20-MeV training range to 92.3\% for the full range.

\item[Conclusions] The high-energy TKE width is particularly sensitive to the incident-energy range represented in the calculated data. The intermediate 30-MeV controls show that the improvement is already evident before the training range reaches 40 MeV. The high-energy GEF width is comparatively slowly varying for the nuclei considered, which limits the extent to which the machine-learning improvement can be attributed to learning a nontrivial high-energy evolution. Adequate incident-energy coverage is therefore critical in the present analysis for reproducing the calculated high-energy evolution of the TKE distribution, while agreement with experiment remains constrained by the underlying fission model.

\end{description}

\end{abstract}

\maketitle

\section{Introduction}\label{sec:intro}

Nuclear fission is a complex collective process in which fragment observables reflect both the evolution of the fissioning system and the configuration reached near scission~\cite{Andreyev2018,SchmidtJurado2018,SchunckRegnier2022}. The total kinetic energy (TKE) carried by the fragments constitutes a major fraction of the prompt fission-energy release and is sensitive to the charge partition and scission configuration. Its dependence on incident-neutron energy, together with the distribution width and spectral profile, provides information that is not contained in the mean TKE alone~\cite{GEF2016,Denisov2022,Shimada2021,Pica2022}. Recent microscopic and dynamical calculations have further emphasized the information carried by TKE distributions, TKE--mass correlations, and fragment kinetic energies, including generalized time-dependent generator-coordinate, Langevin--evaporation, and dynamical scission approaches~\cite{Li2025,Liu2025,Takagi2025}.

At incident energies of several tens of MeV, pre-fission emission and multichance fission increasingly mix contributions from different fissioning systems. The resulting evolution of the TKE centroid, width, and spectral profile therefore provides a demanding high-energy test of fission calculations, precisely in a region where measured distributions are comparatively sparse~\cite{Pica2022}. Dynamical calculations also indicate that kinetic-energy distributions can be sensitive to pre-scission neutron emission and multichance fission~\cite{Ivanyuk2025}, providing additional motivation for examining the energy evolution of the complete TKE distribution rather than its centroid alone.

The General Description of Fission Observables (\gef) provides systematic post-neutron TKE calculations over nuclear systems and incident energies for which experimental distributions are incomplete or unavailable~\cite{GEF2016}. Probabilistic machine-learning models trained on these calculations can provide continuous representations of the dependence on target charge $Z$, mass $A$, incident-neutron energy $E_n$, and post-neutron total kinetic energy $T$. Two distinct questions must then be kept separate: how accurately the learned representation reproduces the calculation on which it was trained, and how the resulting predictions compare with independent measurements. The purpose of the learned representation is not to replace \gef\ as a fission model. Rather, it provides a continuous and computationally inexpensive probabilistic representation of the calculated response in $(Z,A,E_n,T)$, enabling systematic interpolation and extrapolation studies together with model-based predictive intervals. This formulation also allows discrepancies between the learned representation and \gef\ to be distinguished explicitly from discrepancies between the underlying fission calculation and experiment.

Machine-learning methods are increasingly used across nuclear and many-body physics~\cite{Boehnlein2022,Carleo2019}. Gaussian processes and Bayesian neural networks have been used for uncertainty quantification and extrapolation of nuclear observables~\cite{Neufcourt2018,Negoita2019,Lasseri2020}. In fission, Bayesian neural networks were introduced for the evaluation of incomplete energy-dependent fission yields~\cite{Wang2019}, subsequently extended to charge-yield reconstruction and systematic studies of network architecture~\cite{Qiao2021,WangPei2021}, and combined with heterogeneous-data fusion for imperfect and incomplete fission-yield data~\cite{Wang2022}. Mixture-density and related probabilistic approaches have likewise been applied to fission-yield distributions~\cite{Lovell2020,Prassa2024}, while a Bayesian mixture-density network has recently been developed for neutron-induced charge yields, combining variational Bayesian layers with a mixture-density output~\cite{Prassa2025}. A recent physics-embedded BNN further incorporated shell-related nuclear information to predict the energy dependence and fine structure of fission-product yields~\cite{Chen2026}. These developments motivate the present use of probabilistic learned representations for full TKE spectra, for which both extrapolation behavior and predictive intervals can be examined explicitly.

The present analysis considers both a common four-model comparison using GPR, GPR-WK, MDN, and BMDN and an extended neural-network study spanning a broader range of target nuclei and incident energies. Transfer to target nuclei not represented in model fitting is examined explicitly, with $^{235}\mathrm{U}(n,f)$ retained as a dedicated excluded-target evaluation population. The predicted distributions are also compared with 60 measured post-neutron TKE distributions for $^{235}$U, $^{237}$Np, $^{239}$Pu, $^{240}$Pu, and $^{242}$Pu over a range of incident-neutron energies~\cite{Yanez2018,PicaNp2020,Chemey2020,Pica2022}. These experimental comparisons provide an external assessment of the calculated energy dependence of the TKE distributions. \gef\ itself is a global semi-empirical framework constrained by broad experimental systematics~\cite{GEF2016,SchmidtJurado2018}; consequently, agreement of the learned representation with experiment should be interpreted primarily as an assessment of the underlying \gef\ trends rather than as an independent data-driven prediction.

Particular attention is given to the evolution of the TKE-distribution width at higher incident-neutron energies. We examine how the incident-energy range represented in the calculated training data affects the predicted high-energy widths and their associated model-based predictive uncertainties. This provides a controlled test of whether extending the training domain improves the learned representation of the high-energy \gef\ behavior, while comparisons with measured spectra assess whether the corresponding calculated trends remain compatible with experiment.

The paper is organized as follows. Section~\ref{sec:models} defines the probabilistic models and predictive-interval methodology. Section~\ref{sec:method} introduces the evaluation metrics, experimental data, and training-energy evaluation protocol. Section~\ref{sec:comparison} presents the four-model analysis, including its calculated dataset and model setup. Section~\ref{sec:full21} presents the extended-domain analysis, the systematic comparison with measured spectra, and the study of the training-energy range. The implications for high-energy TKE systematics, extrapolation, and predictive uncertainty are discussed in Sec.~\ref{sec:discussion}.

\section{Probabilistic models and uncertainty quantification}\label{sec:models}

\subsection{GEF reference and model inputs}

All systems studied here undergo neutron-induced fission. Throughout this work, a nuclide appearing in the reaction notation ${}^{A}_{Z}X(n,f)$ denotes the target nucleus.

For each target and incident-neutron energy, \gef\ provides the post-neutron fragment TKE distribution as calculated counts at successive TKE values. The four model inputs and predicted quantity are

\begin{equation}
 \bm{x}=(Z,A,E_n,T),\qquad
 Y=Y_{\mathrm{GEF}}(Z,A,E_n,T),
 \label{eq:target}
\end{equation}

where $T$ denotes the post-neutron TKE and $Y$ the corresponding value from the GEF spectrum. A full TKE distribution is reconstructed by evaluating the model over successive values of \(T\) at fixed \((Z,A,E_n)\). The GEF spectra consist of calculated spectral counts on a common scale of approximately \(10^5\) counts per spectrum, with the integrated post-neutron counts differing from this nominal value by less than 0.4\%. Pointwise errors can therefore be compared on a consistent scale. Because the GEF spectra are obtained from finite Monte-Carlo samples, the pointwise error and coverage measures include small statistical fluctuations of the reference spectra. Spectrum-level observables such as the mean TKE and distribution width are less sensitive to these bin-to-bin fluctuations.

All \gef{} calculations used in this work were generated with version 2023/3.3 using the global parameter set. The same setup is used for the four-model and extended-domain training data, the excluded-target calculations, and the direct comparisons with experiment. The analysis uses the post-neutron \texttt{Counts(post)} distribution from the GEF \texttt{<TKE>} output. At incident energies where additional fission chances become accessible, GEF accounts for multichance fission, including neutron emission prior to fission.

\subsection{Gaussian-process models}

GPR represents the dependence of $Y$ on the four inputs through a Gaussian process. The predictive mean estimates the calculated \gef\ count at each TKE value, while the predictive variance expresses the uncertainty assigned by the GP at the specified input~\cite{GPML}. The two implementations differ in their covariance functions:

\begin{align}
 k_{\mathrm{GPR}}(\bm x,\bm x') &= k_{\mathrm{M52}}(\bm x,\bm x')
 +k_{\mathrm{RBF}}(\bm x,\bm x'),\\
 k_{\mathrm{GPR\mbox{-}WK}}(\bm x,\bm x') &= k_{\mathrm{GPR}}(\bm x,\bm x')
 +k_{\mathrm{white}}(\bm x,\bm x').
\end{align}

The Mat\'ern-5/2 and radial-basis (RBF) terms describe correlations across the input domain. GPR-WK additionally represents an uncorrelated contribution to the modeled spectral values through a white-noise kernel; that term is not a measurement of physical fluctuations in the fission process.  Both implementations return a predictive Gaussian distribution on the original \gef\ count scale.

The GPR and GPR-WK kernel settings are selected using only the calculated training data. Complete $(Z,A,E_n)$ spectra are kept together when constructing three independent grouped training--validation partitions, so that TKE channels from the same spectrum do not appear simultaneously in the fitting and validation subsets. For each partition, the input and target \texttt{RobustScaler} transformations are fitted on the fitting subset and then applied to the corresponding validation subset.

Kernel candidates are ranked using the Gaussian negative log predictive density (NLPD),
\begin{equation}
\mathrm{NLPD}
=
\frac{1}{N}\sum_{i=1}^{N}
\left[
\frac{1}{2}\ln\!\left(2\pi\sigma_i^2\right)
+
\frac{(y_i-\mu_i)^2}{2\sigma_i^2}
\right],
\label{eq:nlpd}
\end{equation}
where $\mu_i$ and $\sigma_i^2$ are the predictive mean and variance for validation point $i$. Unlike an error measure based only on the predictive mean, NLPD evaluates the full predictive distribution and penalizes both inaccurate predictions and predictive variances that are inconsistent with the observed residuals~\cite{GPML,GneitingRaftery2007}. The mean NLPD across the three grouped partitions is used for kernel selection; empirical coverage, RMSE, and predictive-interval width are retained as diagnostics and are not used as selection criteria. The independent calculated test set, the excluded $^{235}$U spectra, and all experimental data are excluded from hyperparameter selection.

The selected scaled-input length scales are $1.915546$ (Mat\'ern) and $0.546119$ (RBF) for GPR, and $1.418687$ (Mat\'ern) and $0.735882$ (RBF) for GPR-WK. The selected GPR-WK white-kernel variance is $0.01$ in scaled target units; both models use a small numerical stabilization term, $\epsilon_{\rm GP}=10^{-6}$. After kernel selection, the preprocessing transformations are fitted on the complete 4,499-point training set and the covariance models are refitted using all training points.

For these stationary Mat\'ern/RBF kernels, the learned contribution to the GP predictive mean at a new point has the form $\bm{k}_{*}^{\mathsf T}(K+\epsilon_{\rm GP}I)^{-1}(\bm{y}-\bm{m}_0)$, where $K$ is the covariance matrix generated by the corresponding kernel and $\bm{k}_{*}$ contains covariances between the new input and the training inputs. For GPR-WK, $K$ additionally contains the contribution of the white-noise kernel. As covariance with the fitted region decreases along an unrepresented energy or nuclear-system direction, the predictive mean approaches the model's prior baseline in the scaled target space~\cite{GPML,PhysicsInformedGP2024}. The high-energy Pu results therefore characterize the extrapolation behavior of the selected stationary kernels; alternative spectral-mixture, nonstationary, physics-guided, or sparse kernels may behave differently~\cite{WilsonAdams2013,VargasHernandez2018,DaiKrems2020}.

Exact GPR is used for the four-model comparison. Its covariance matrix contains \(N\times N\) elements for \(N\) training points, and standard factorization requires approximately \(\mathcal O(N^3)\) operations and \(\mathcal O(N^2)\) storage. Consequently, the computational cost increases rapidly as the number of training points grows, making exact GPR impractical for the substantially larger extended dataset. Inducing-point and stochastic-variational GP approximations provide established routes to improved scaling~\cite{Titsias2009,Hensman2013}; because such approximations are not used here, the extended-domain analysis focuses on the neural probabilistic models.

\subsection{Mixture-density network}

Unlike a single-output regression, the MDN assigns a conditional probability distribution to the spectral value $Y$ at a given input~\cite{Bishop1994}. Mixture-density networks have previously been applied to fission mass-yield distributions and their uncertainties~\cite{Prassa2024,Lovell2020}.

The comparative-study architecture contains four inputs and one fully connected hidden layer with 16 sigmoid units. Its output consists of three Gaussian components in the transformed variable $u=\log(1+Y)$:

\begin{equation}\label{eq:mdn}
 p(u\mid\bm x)=\sum_{k=1}^{3}\pi_k(\bm x)\,
 \mathcal N\!\left(u\mid\mu_k(\bm x),s_k^2(\bm x)\right).
\end{equation}

The network predicts each component's relative weight $\pi_k$ (normalized with a softmax), mean $\mu_k$, and positive standard deviation $s_k$ (obtained with a softplus transformation and a small positive offset). The model is trained by minimizing the negative log likelihood of the calculated \gef\ response. Predictions and intervals are then converted back to the original spectral-value scale, $Y=\exp(u)-1$. Because the mixture is Gaussian in $u=\log(1+Y)$, the conditional mean used as the central MDN prediction on the original scale is

\begin{equation}\label{eq:mdnmean}
\widehat{Y}(\bm x)=\sum_{k=1}^{3}\pi_k(\bm x)
\left[\exp\!\left(\mu_k(\bm x)+\frac{s_k^2(\bm x)}{2}\right)-1\right].
\end{equation}

The three Gaussian components provide a flexible representation of the conditional response \(p(u\mid\mathbf{x})\). They are statistical mixture components and are not associated with distinct physical fission modes. Spectrum-level moments are evaluated from the conditional-mean spectrum of Eq.~(\ref{eq:mdnmean}), consistently with the pointwise central prediction used throughout the reconstruction analysis.

\subsection{Bayesian mixture-density network}

The BMDN retains the same three-component conditional mixture as the MDN but replaces the deterministic hidden layer with a 16-unit variational sigmoid layer. The hidden-layer weights are represented by an approximate posterior \(q(\mathbf{w})\) rather than by a single fitted set of values, following the variational Bayesian neural-network framework of Ref.~\cite{Blundell2015}. For each sampled weight realization, the conditional mean is evaluated using the same mixture expression as in Eq.~(\ref{eq:mdnmean}), and the reported BMDN central prediction is obtained by averaging these conditional means over the variational weight samples.

In the implemented \texttt{DenseVariational} layer, the Gaussian weight prior and variational posterior are parameterized as trainable full-covariance multivariate normal distributions. The objective combines the mixture negative log likelihood with a Kullback--Leibler (KL) contribution weighted by \(10^{-5}\). The prior and posterior are optimized jointly during training, and sampling from the resulting weight distribution provides the variational contribution to the BMDN predictive uncertainty.

Sampling the network weights changes the conditional mixture, so the BMDN represents both the conditional spread at fixed weights and the variation of the predicted mean across the variational weight distribution.

Only the hidden \texttt{DenseVariational} layer is treated variationally; the output layers that map the hidden representation to mixture weights, component means, and component scales are deterministic. The variational-weight contribution defined below therefore reflects the variation induced by the variational hidden-layer weights. Sensitivity to the prior specification is examined by repeating the BMDN calculations with a fixed standard-normal prior, \(p(\mathbf w)=\mathcal N(\mathbf 0,\mathbf I)\), while retaining the same architecture, variational posterior, KL weighting, preprocessing, and training protocol. The same prior-only replacement is applied to the extended-domain training-range analysis across the three training ranges, three grouped partitions, and three neural initializations.

The two contributions to its total predictive variance follow the law of total variance:

\begin{align}
\label{eq:decomp}
\underbrace{\mathrm{Var}(Y\mid\bm x,\mathcal D)}_{V_{\mathrm{tot}}}
={}&\underbrace{\mathbb E_{q(\bm w)}\!\left[
 \mathrm{Var}(Y\mid\bm x,\bm w)\right]}_{V_{\mathrm{mix}}}\\
 &+\underbrace{\mathrm{Var}_{q(\bm w)}\!\left[
 \mathbb E(Y\mid\bm x,\bm w)\right]}_{V_{\mathrm{w}}}.\nonumber
\end{align}

A related predictive-variance decomposition is commonly used in Bayesian neural-network uncertainty analyses~\cite{KendallGal2017}. Here \(V_{\mathrm{mix}}\) is the conditional-mixture variance averaged over variational weight samples, and \(V_{\mathrm{w}}\) is the variance of the predicted mean across variational weight samples. Both are evaluated on the original GEF spectral-value scale. This is a model-based variance partition; neither term is interpreted as experimental uncertainty or as uncertainty in the underlying GEF physics. Their relative contributions are reported as \(100V_{\mathrm{mix}}/V_{\mathrm{tot}}\) and \(100V_{\mathrm{w}}/V_{\mathrm{tot}}\).

\subsection{Predictive intervals}\label{sec:training}

The 90\% and 95\% shaded regions in comparisons on the GEF count scale show pointwise predictive intervals (PIs) for the calculated spectral value at each sampled TKE. In the four-model comparison, MDN intervals are estimated from 4,000 samples drawn from the learned mixture, whereas BMDN intervals use 200 variational-weight realizations with 20 conditional-mixture samples per realization, also giving 4,000 predictive samples per point. In the extended-domain analysis, MDN intervals are estimated from 1,024 mixture samples per point, while BMDN uses 128 variational-weight realizations with eight conditional-mixture samples per realization, again giving 1,024 predictive samples per point. For GPR and GPR-WK, the count-scale intervals are obtained from the fitted Gaussian predictive distributions; the GPR-WK predictive covariance includes the \texttt{WhiteKernel} contribution. Independent neural fits are reported separately to assess training reproducibility.

For comparisons with experimental spectral shapes, predictive uncertainty is propagated using complete spectral realizations. For each realization  $Y^{(s)}(T)$, negative ordinates are projected to zero and the spectrum is normalized by its trapezoidal area,

\begin{equation}
 p^{(s)}(T)=\frac{\max[Y^{(s)}(T),0]}{\int dT\,\max[Y^{(s)}(T),0]}.
 \label{eq:shapePI}
\end{equation}

The pointwise 90\% and 95\% predictive intervals of the normalized spectral shape are obtained from the corresponding quantiles of \(p^{(s)}(T)\). Because the normalization depends on the complete spectrum, these intervals are constructed from full spectral realizations rather than independently at each TKE point. For GPR and GPR-WK, complete spectra are sampled jointly from the multivariate GP predictive distribution over the TKE grid, preserving correlations among TKE points. For MDN and BMDN, complete realizations are sampled from their respective predictive distributions; for BMDN, each sampled weight realization is retained across the full TKE grid before sampling from the conditional mixture. The central prediction is treated in the same way by enforcing nonnegativity and normalizing to unit area. When the distributions are displayed on the experimental ordinate scale, normalization to the experimental area is applied only after the normalized-shape statistics have been constructed.

Predictive-interval performance is assessed using both empirical coverage and the interval score. For a nominal \((1-\alpha)\) predictive interval \([l,u]\) and reference value \(y\), the interval score is~\cite{GneitingRaftery2007}

\begin{equation}\label{eq:intervalscore}
\mathrm{IS}_{\alpha}(l,u;y)
=(u-l)+\frac{2}{\alpha}(l-y)_{+}
+\frac{2}{\alpha}(y-u)_{+},
\end{equation}

where \((a)_{+}=\max(a,0)\). Smaller values indicate a better balance between interval width and missed reference values. We use \(\alpha=0.05\) and average the score over the same evaluation points used for coverage. 

\section{Data and evaluation}\label{sec:method}

\subsection{Evaluation metrics}
\label{sec:metrics}

The same pointwise and spectrum-level evaluation measures are used in the four-model and extended-domain analyses. Two levels of comparison are considered. Pointwise measures quantify reconstruction of the GEF spectral values at the sampled TKE points, whereas spectrum-level quantities characterize each complete TKE distribution through its mean, width, and normalized spectral profile.

For an evaluation population containing \(N_{\rm pt}\) spectral points,
\begin{align}
{\rm RMSE}_{\rm pt}
&=
\left[
\frac{1}{N_{\rm pt}}
\sum_{\ell=1}^{N_{\rm pt}}
\left(\widehat Y_\ell-Y_\ell^{\rm GEF}\right)^2
\right]^{1/2},
\label{eq:point_rmse}\\
{\rm MAE}_{\rm pt}
&=
\frac{1}{N_{\rm pt}}
\sum_{\ell=1}^{N_{\rm pt}}
\left|\widehat Y_\ell-Y_\ell^{\rm GEF}\right|.
\label{eq:point_mae}
\end{align}
Both quantities are reported in GEF count units.

For a calculated or predicted spectrum, let \(c_j\) denote the spectral value at TKE \(T_j\). For spectrum-level observables, negative fitted values are set to zero, \(c_j=\max(0,\widehat Y_j)\). Pointwise reconstruction measures and count-scale predictive-interval diagnostics are evaluated without applying the nonnegativity projection. Normalized spectral-shape intervals are constructed as described in Sec.~\ref{sec:training}.

With \(p_j=c_j/\sum_k c_k\), the mean and width are
\begin{equation}
\label{eq:meanwidth}
\langle T\rangle=\sum_j p_jT_j,\qquad
\sigma_T=\left[\sum_j p_j(T_j-\langle T\rangle)^2\right]^{1/2}.
\end{equation}

For a set \(\mathcal S\) of \(N_{\mathcal S}\) complete spectra, the mean absolute error of either spectrum-level observable \(q\in\{\langle T\rangle,\sigma_T\}\) is
\begin{equation}
{\rm MAE}_{q}(\mathcal S)
=
\frac{1}{N_{\mathcal S}}
\sum_{i\in\mathcal S}
\left|q_i^{\rm calc}-q_i^{\rm ref}\right|,
\label{eq:spectrum_mae}
\end{equation}
where the reference is GEF for calculated-data validation and experiment for external comparisons. Neural-model quantities are evaluated separately for each independent fit or training configuration before being summarized across fits or configurations as specified in the corresponding table captions.

Normalized spectral profiles are additionally examined to retain information on peak position, broadening, and shoulders. Predictive-interval performance is assessed through coverage, interval width, and interval score, as described in Sec.~\ref{sec:training}.

GEF mean-TKE and width values are calculated directly from the post-neutron GEF TKE spectrum at the representative incident energy. Interpolation onto experimental or model TKE coordinates is used only when a common grid is required for pointwise spectral comparisons or for the explicitly identified second-moment sensitivity analysis.

\subsection{Experimental data}\label{sec:exp60}\label{sec:pudata}
\begin{table*}[!tb]

\caption{Experimental post-neutron TKE distributions used for comparison with the calculations and model predictions. The listed values are the representative incident-neutron-energy group centres used for the \gef\ calculations and ML queries. The experimental groups have finite energy widths; the calculations are evaluated at the representative energies listed here.}

\label{tab:exp60coverage}\centering

\begingroup\small\setlength{\tabcolsep}{5.5pt}\renewcommand{\arraystretch}{1.16}

\begin{tabular}{lcp{0.76\textwidth}}\toprule
Target & $N$ & Representative $E_n$ values (MeV)\\
\midrule
$^{235}$U & 20 & 2.36, 2.71, 3.11, 3.62, 4.28, 5.08, 6.14, 7.28, 8.54, 10.19, 12.57, 15.72, 19.93, 25.09, 31.45, 38.90, 47.35, 57.73, 71.45, 89.71\\
$^{237}$Np & 10 & 2.96, 3.83, 5.19, 7.06, 9.70, 14.80, 23.90, 37.10, 56.50, 83.80\\
$^{239}$Pu & 10 & 2.94, 3.77, 4.98, 6.73, 9.07, 13.54, 21.82, 34.93, 55.04, 82.66\\
$^{240}$Pu & 10 & 0.78, 1.25, 1.82, 2.61, 3.84, 6.00, 9.51, 17.94, 36.27, 73.36\\
$^{242}$Pu & 10 & 0.85, 1.32, 1.91, 2.77, 4.19, 6.73, 11.16, 21.92, 42.50, 76.55\\
\bottomrule
\end{tabular}\endgroup

\end{table*}

Comparison with experiment uses 60 measured post-neutron TKE distributions from four studies: 20 spectra for \(^{235}\mathrm{U}\)~\cite{Yanez2018}, 10 for \(^{237}\mathrm{Np}\)~\cite{PicaNp2020}, 10 for \(^{239}\mathrm{Pu}\)~\cite{Chemey2020}, and 10 each for \(^{240}\mathrm{Pu}\) and \(^{242}\mathrm{Pu}\)~\cite{Pica2022}. These data are not used for model fitting or selection. The experimental neutron-energy group centres are used as the representative incident energies for the \gef\ calculations and ML predictions; the full energy coverage is listed in Table~\ref{tab:exp60coverage}.

The primary comparison uses the mean post-neutron TKE values and TKE variances reported in the experimental studies. The reported variances are expressed as widths through \(\sigma_T=\sqrt{\sigma_T^2}\). For \gef\ and the ML predictions, the corresponding mean TKE and width are calculated directly from the normalized TKE distributions using Eq.~(\ref{eq:meanwidth}).

To assess sensitivity to the width definition, the experimental widths are also recalculated from the measured TKE distributions using the same second-moment expression already used for the \gef\ and model predictions. These recalculated experimental widths are used only for the explicitly identified width-definition sensitivity analysis.

For spectral-shape comparisons, the \gef\ and central model predictions are evaluated on the experimental TKE coordinates and normalized independently. Predictive intervals for the normalized model spectra are constructed realization by realization according to Eq.~(\ref{eq:shapePI}). In the four-model comparison, representative \(^{240}\mathrm{Pu}(n,f)\) spectra at 17.94 and 36.27 MeV are used to illustrate behavior within and beyond the original 20-MeV training range. In the extended-domain analysis, representative high-energy comparisons are shown for \(^{237}\mathrm{Np}(n,f)\) at 56.50 MeV~\cite{PicaNp2020} and \(^{240}\mathrm{Pu}(n,f)\) at 73.36 MeV~\cite{Pica2022}.

\subsection{Training-energy controls and evaluation protocol}\label{sec:ablationmethod}

The dependence on the training-energy range is studied with MDN and BMDN models trained on calculated spectra restricted to \(E_n\le20\) MeV, \(E_n\le40\) MeV, or the full available range extending to approximately 76 MeV. The model architecture and training procedure are unchanged among these cases, and the same grouped partitions and neural initializations are used throughout. For each energy cutoff, both the fitting and validation populations are restricted consistently so that spectra above the specified range enter neither parameter estimation nor model selection. For BMDN, the KL weighting follows the same prescription in all cases. Sensitivity to the prior specification is examined separately by repeating the BMDN calculations with a fixed standard-normal prior, \(p(\mathbf w)=\mathcal N(\mathbf 0,\mathbf I)\), while keeping the remaining settings unchanged.

Because extending the energy range also increases the number of available calculated spectra, matched-size controls are constructed for upper limits of 30 and 40 MeV. For each target and grouped partition, these controls retain the same numbers of complete fitting and validation spectra as the \(E_n\le20\) MeV population while replacing part of the lower-energy sample with spectra from the corresponding higher-energy interval. Two independently predetermined \(E_n\le30\) MeV realizations are considered to assess sensitivity to the particular matched subset. These controls are used to distinguish the effect of broader incident-energy coverage from that of increasing the size of the training population.

All reconstruction and spectrum-level quantities are evaluated separately for each combination of grouped partition and neural initialization before averages across the resulting nine training configurations are formed. Variation among these configurations is used to characterize training reproducibility and is distinguished from the predictive uncertainty of an individual fitted model. For BMDN, the latter is further separated into conditional-mixture and variational-weight contributions.

The principal training-range comparison focuses on the experimental high-energy width errors. Differences between training conditions are assessed with paired bootstrap resampling over training configurations and experimental spectra. A complementary target-cluster bootstrap resamples the five target nuclei to examine the stability of the result across nuclear systems. The matched \(E_n\le30\) MeV controls are evaluated on the common \(E_n>30\) MeV experimental subset, whereas the \(E_n>40\) MeV subset is used for the main high-energy comparison. Sensitivity to individual grouped partitions and neural initializations is additionally examined through leave-one-out checks.

Two target-specific diagnostics are used to characterize the high-energy \gef\ width trend independently of the neural models. A linear continuation and a constant-width carry-forward are constructed from the \gef\ widths available at \(E_n\le20\) MeV and evaluated at the same high-energy experimental energies. They are used only to characterize the calculated energy dependence and not as alternative predictive models.

Finally, the uncertainties of the reported experimental TKE variances are propagated to the corresponding widths and included in the paired bootstrap to test the stability of the high-energy training-range result.

\section{Four-model comparison}\label{sec:comparison}

The four-model analysis is used first to compare the behavior of GPR, GPR-WK, MDN, and BMDN under the same calculated training domain. The comparison is organized in three steps: reconstruction of held-out GEF spectra, probabilistic-interval behavior, and transfer to a target nucleus excluded from training. Experimental \(^{240}\mathrm{Pu}(n,f)\) spectra are then used only as illustrative external cases and to motivate the broader energy-domain study of Sec.~\ref{sec:full21}.

\subsection{Dataset and model setup}

The four-model comparison uses 4,499 spectral points from 44 complete GEF TKE distributions. Model development is based on spectrum-grouped fitting and validation subsets containing 3,503 and 996 points, respectively, with complete \((Z,A,E_n)\) spectra assigned to a single subset. After model settings are selected, the final models are fitted using the complete 4,499-point training population. An independent test set contains 1,117 points from 11 complete TKE distributions. Five additional \(^{235}\mathrm{U}(n,f)\) distributions, comprising 455 spectral points, are excluded from both fitting and model selection and are used as a separate transfer test. The corresponding population sizes are summarized in Table~\ref{tab:sets}. The target nuclei and incident-neutron energies represented in the training population are listed in Table~\ref{tab:fourcoverage}. The independent test spectra lie within the overall incident-energy range represented in training.

The MDN and BMDN use the same four input variables and scaling procedure. Each neural model is trained in three independent runs using Adam~\cite{KingmaBa2015} with an initial learning rate of \(3\times10^{-3}\) and a batch size of 32. Hyperparameters and training settings are selected using the spectrum-grouped fitting and validation data summarized in Table~\ref{tab:sets} and are fixed before evaluation on the independent test set and the excluded \(^{235}\mathrm{U}(n,f)\) spectra.

\begin{table}[!t]

\caption{Data populations used in the four-model comparison. The fitting and validation subsets are used for model development; after model selection, the complete 4,499-point training set is used for the final fits. The independent test set and all \(^{235}\mathrm{U}(n,f)\) spectra are excluded from model selection.}

\label{tab:sets}

\centering

\begingroup

\setlength{\tabcolsep}{10pt}

\renewcommand{\arraystretch}{1.09}

\begin{tabular}{lrr}
\toprule
Population & Data points & Spectra\\
\midrule
Training set & 4,499 & 44\\
\quad Fitting subset & 3,503 & 35\\
\quad Validation subset & 996 & 9\\
Independent test set & 1,117 & 11\\
\(^{235}\mathrm{U}(n,f)\) & 455 & 5\\
\bottomrule
\end{tabular}

\endgroup

\end{table}

\begin{table}[!t]

\caption{Target nuclei and incident-neutron energies \(E_n\) (MeV) represented in the 4,499-point training dataset used for the four-model comparison. Mass numbers refer to the target nuclei.}

\label{tab:fourcoverage}

\centering

\begingroup

\setlength{\tabcolsep}{12pt}

\renewcommand{\arraystretch}{1.09}

\begin{tabular}{ll}
\toprule
Target & \(E_n\) (MeV) in training set \\
\midrule
\(^{229}\mathrm{Th}\) & \(2.53\times10^{-8}\), 10, 15 \\
\(^{230}\mathrm{Pa}\) & \(2.53\times10^{-8}\), 10, 15, 20 \\
\(^{234}\mathrm{U}\) & \(2.53\times10^{-8}\), 5, 10, 15, 20 \\
\(^{236}\mathrm{U}\) & \(2.53\times10^{-8}\), 5, 15, 20 \\
\(^{237}\mathrm{Np}\) & 5, 10, 15 \\
\(^{239}\mathrm{Pu}\) & \(2.53\times10^{-8}\), 5, 10, 15, 20 \\
\(^{243}\mathrm{Am}\) & \(2.53\times10^{-8}\), 10, 20 \\
\(^{248}\mathrm{Cm}\) & \(2.53\times10^{-8}\), 5, 10, 20 \\
\(^{251}\mathrm{Cf}\) & \(2.53\times10^{-8}\), 5, 10, 15 \\
\(^{253}\mathrm{Es}\) & \(2.53\times10^{-8}\), 5, 10, 20 \\
\(^{255}\mathrm{Fm}\) & \(2.53\times10^{-8}\), 5, 10, 15, 20 \\
\bottomrule
\end{tabular}

\endgroup

\end{table}

 \begin{figure*}[t]
\centering
\includegraphics[width=.98\textwidth]{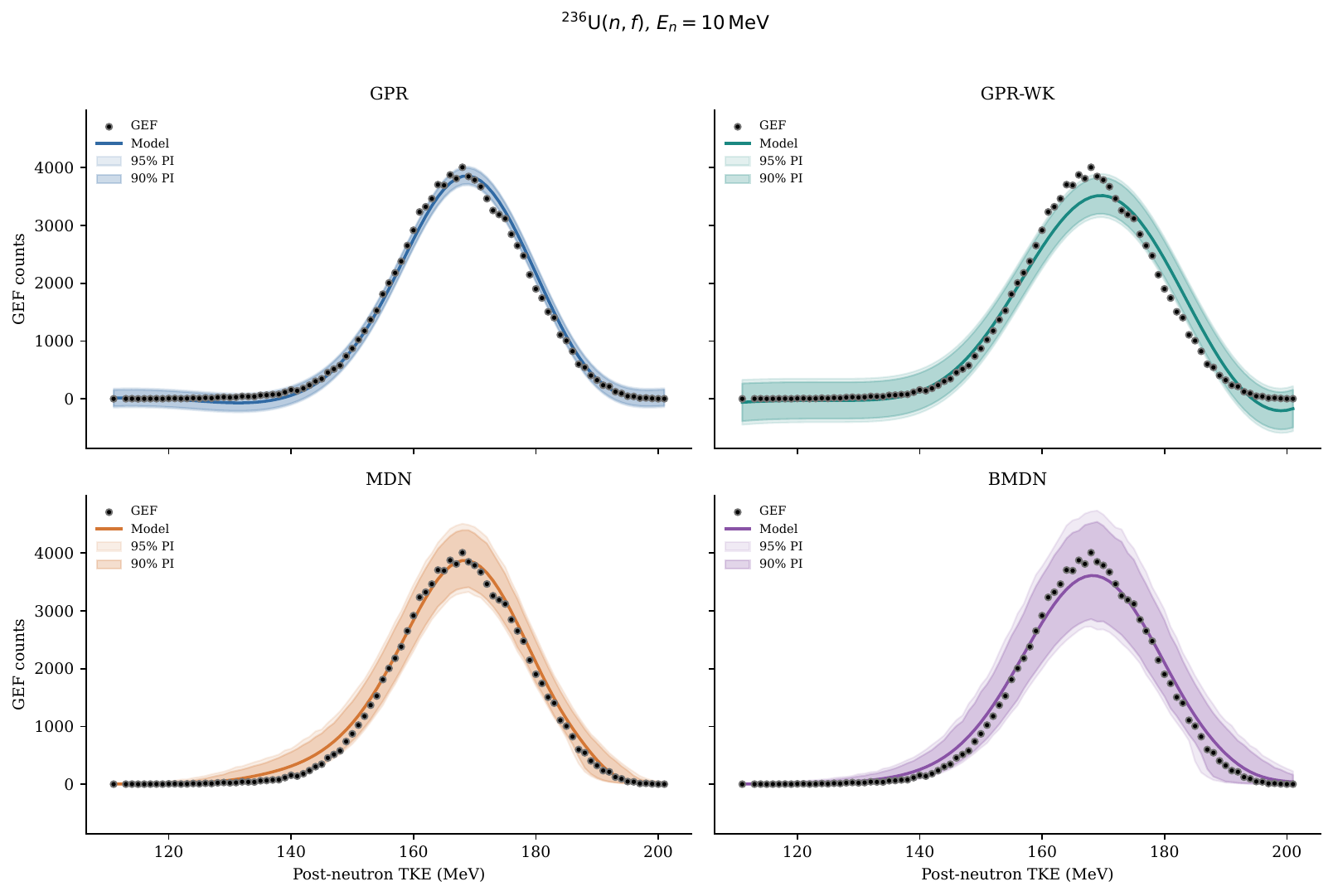}
\caption{Post-neutron TKE distribution for \(^{236}\mathrm{U}(n,f)\) at \(E_n=10\) MeV from the independent test set, shown for the four models. Filled black markers show the GEF spectral values, solid colored curves the model predictions, and shaded regions the 90\% and 95\% pointwise predictive intervals.}
\label{fig:test}
\end{figure*}

Figure~\ref{fig:test} shows a representative \(^{236}\mathrm{U}(n,f)\) spectrum from the independent test set at \(E_n=10\) MeV. All four models reproduce the main bell-shaped structure of the GEF distribution.

\subsection{Reconstruction of held-out GEF spectra}

Pointwise reconstruction accuracy is evaluated on an independent test set comprising 11 complete GEF TKE distributions from eight target nuclei, all lying within the overall incident-energy range represented in the training data. MDN gives the lowest RMSE, \(157.62\pm38.38\) counts, followed by BMDN with \(169.59\pm21.50\) counts and GPR with 176.70 counts, while GPR-WK gives the largest value, 216.35 counts (Table~\ref{tab:acc}). The corresponding MAEs show the same overall ordering.

\begin{table}[!t]
\caption{Pointwise reconstruction errors relative to GEF for the 11 independent test spectra. RMSE and MAE are defined in Eqs.~(\ref{eq:point_rmse}) and (\ref{eq:point_mae}) and are reported in count units. MDN and BMDN values are the mean \(\pm\) sample standard deviation over three independent fits; each GP model uses a single fit.}
\label{tab:acc}
\centering
\begingroup
\setlength{\tabcolsep}{16pt}
\renewcommand{\arraystretch}{1.13}
\begin{tabular}{lrr}
\toprule
Model & RMSE & MAE\\
\midrule
GPR & 176.70 & 115.75\\
GPR-WK & 216.35 & 147.55\\
MDN & $157.62\pm38.38$ & $96.14\pm19.48$\\
BMDN & $169.59\pm21.50$ & $101.23\pm8.41$\\
\bottomrule
\end{tabular}
\endgroup
\end{table}

The same test spectra are also compared through the spectrum-level observables of Eq.~(\ref{eq:meanwidth}). Their mean absolute errors relative to GEF are calculated with Eq.~(\ref{eq:spectrum_mae}) and are shown in Table~\ref{tab:physicaltest}. The ranking is not identical to the pointwise ranking: BMDN gives the smallest mean-TKE error, whereas MDN gives the smallest width error. Pointwise errors weight discrepancies at individual TKE values, while the centroid and width depend on the complete normalized distribution.

\begin{table}[!t]
\caption{Spectrum-level mean absolute errors relative to GEF for the 11 independent test spectra. Mean-TKE and width errors are calculated using Eq.~(\ref{eq:spectrum_mae}). MDN and BMDN values are the mean \(\pm\) sample standard deviation over three independent fits; each GP model uses a single final fit. All values are in MeV.}
\label{tab:physicaltest}
\centering
\begingroup
\setlength{\tabcolsep}{12pt}
\renewcommand{\arraystretch}{1.12}
\begin{tabular}{lrr}
\toprule
Model & Mean-TKE MAE & Width MAE\\
\midrule
GPR & 0.981 & 0.624\\
GPR-WK & 0.927 & 0.746\\
MDN & $0.922\pm0.132$ & $0.539\pm0.102$\\
BMDN & $0.844\pm0.034$ & $0.630\pm0.039$\\
\bottomrule
\end{tabular}
\endgroup
\end{table}

\subsection{Predictive intervals and training reproducibility}

The four models also differ in the behavior of their predictive intervals. At the nominal 95\% level, GPR and GPR-WK cover 88.5\% and 93.7\% of the held-out GEF spectral values, respectively, compared with 95.9\% for MDN and 98.7\% for BMDN (Table~\ref{tab:coverage}). Coverage is reported together with mean interval width because high coverage can be obtained simply by broadening an interval; the two quantities therefore provide complementary information on calibration and sharpness.

\begin{table}[t]
\caption{Empirical coverage and mean width of the nominal 95\% pointwise predictive intervals relative to GEF for the 11 independent test spectra. Interval width is reported in count units. MDN and BMDN values are the mean \(\pm\) sample standard deviation over three independent fits; each GP model uses a single fit.}
\label{tab:coverage}
\centering
\begingroup
\setlength{\tabcolsep}{7pt}
\renewcommand{\arraystretch}{1.13}
\begin{tabular}{@{}lrr@{}}
\toprule
Model & Coverage (\%) & Mean interval width\\
\midrule
GPR & 88.54 & 655.59\\
GPR-WK & 93.73 & 870.02\\
MDN & $95.91\pm1.69$ & $542.79\pm169.37$\\
BMDN & $98.69\pm0.37$ & $682.11\pm47.40$\\
\bottomrule
\end{tabular}
\endgroup
\end{table}

\begin{figure*}[!t]
\centering
\includegraphics[width=.95\textwidth]{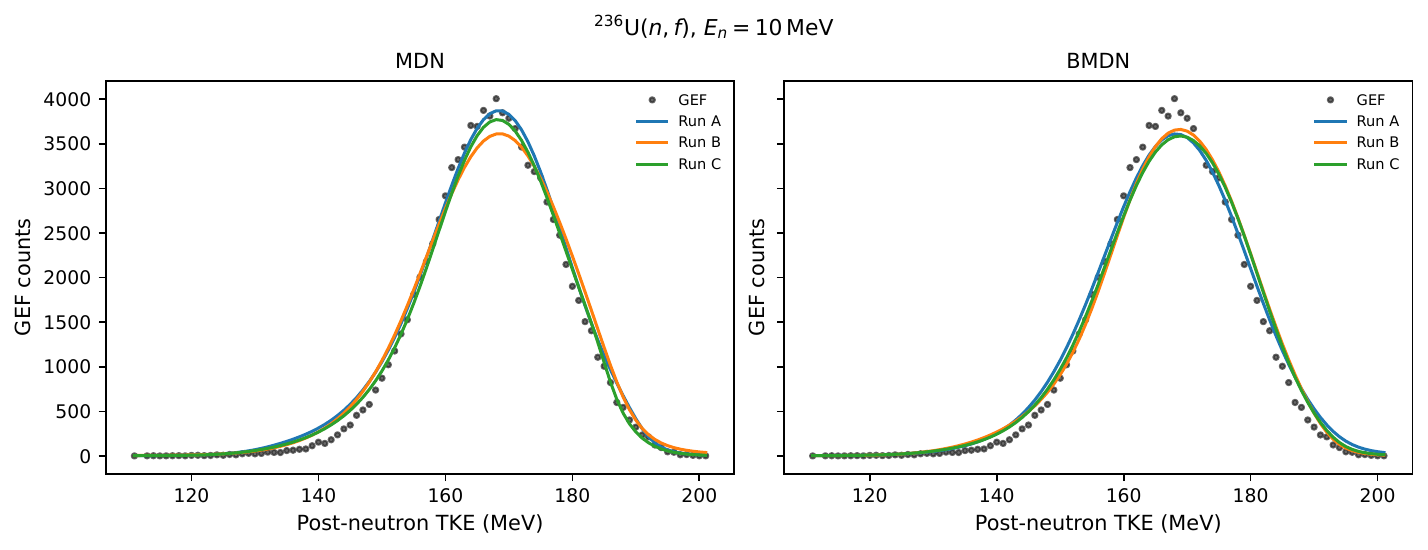}
\caption{Post-neutron TKE distribution for \(^{236}\mathrm{U}(n,f)\) from the independent test set, showing predictions from three independently trained MDN and BMDN models. The spread among the curves illustrates training reproducibility and is distinct from a within-fit predictive interval.}
\label{fig:repro}
\end{figure*}

Training reproducibility for the neural models is illustrated in Fig.~\ref{fig:repro}, which compares three independently trained MDN and BMDN predictions for the same test spectrum. This between-fit variation is distinct from the predictive interval of an individual fitted model.

For BMDN, the conditional-mixture term is the dominant contribution to the modeled predictive variance, accounting on average for 92.0\% of the variance on the test set and 93.9\% for the excluded \(^{235}\mathrm{U}(n,f)\) spectra. Repeating the calculation with a fixed standard-normal prior produces only small changes in the corresponding variance fractions and coverage values.

\begin{figure*}[!t]
\centering
\includegraphics[width=.97\textwidth]{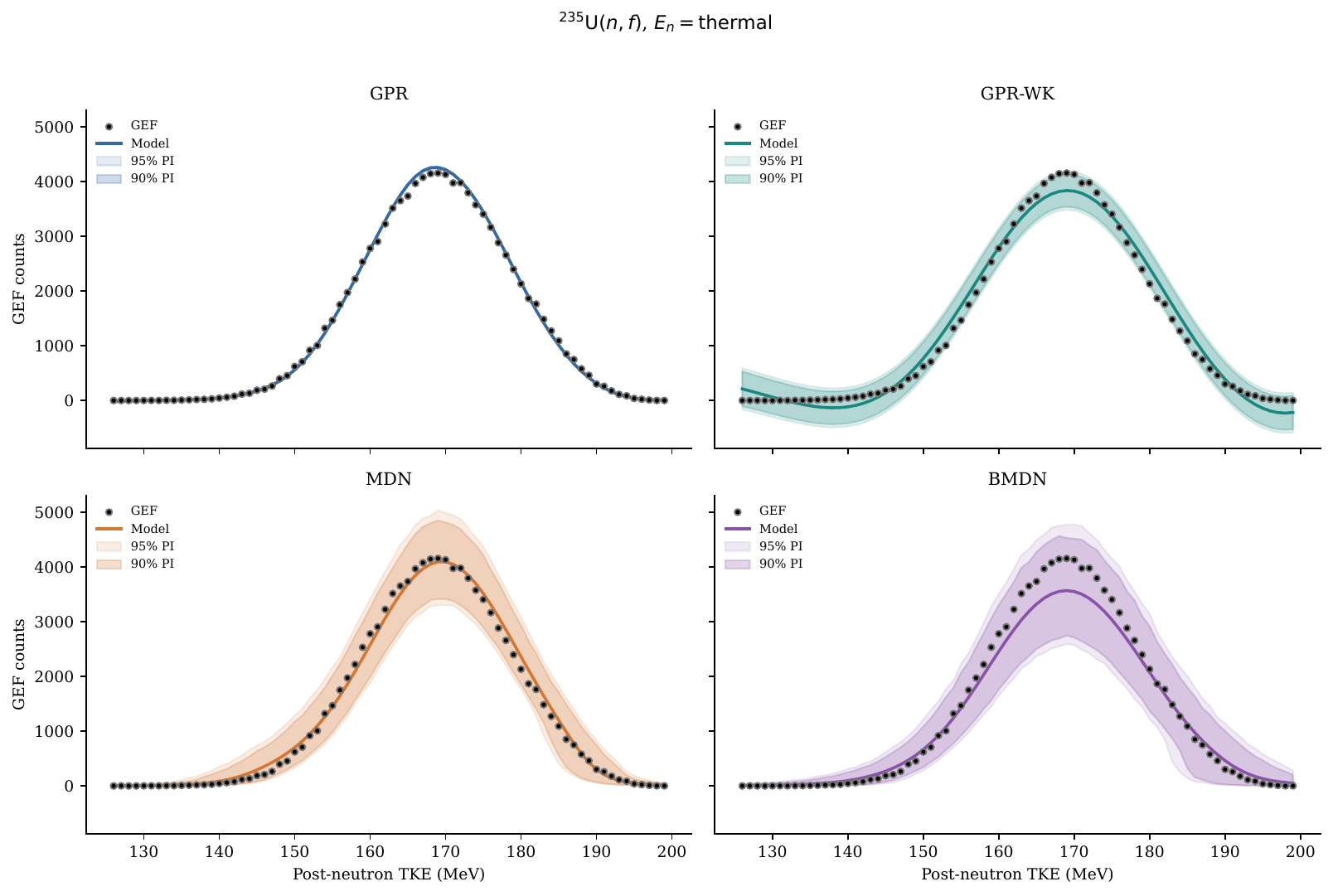}
\caption{Post-neutron TKE distribution for \(^{235}\mathrm{U}(n,f)\) at thermal incident energy. The target nucleus was excluded from training and model selection. Filled black markers show the GEF spectral values, solid curves the model predictions, and shaded regions the 90\% and 95\% pointwise predictive intervals.}
\label{fig:uthermal}
\end{figure*}

\begin{figure*}[!t]
\centering
\includegraphics[width=.97\textwidth]{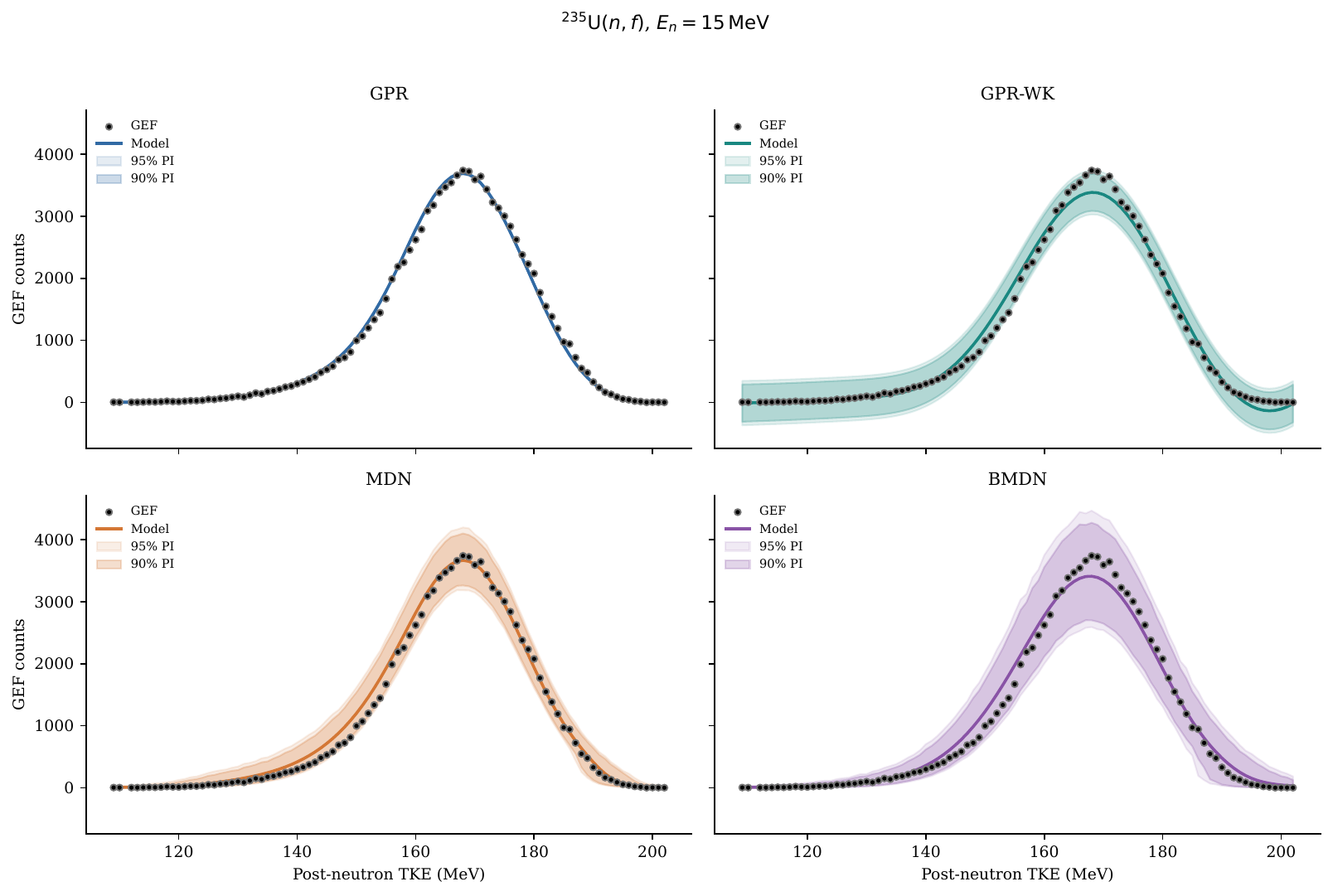}
\caption{Post-neutron TKE distribution for \(^{235}\mathrm{U}(n,f)\) at \(E_n=15\) MeV, with the same conventions as Fig.~\ref{fig:uthermal}.}
\label{fig:u15}
\end{figure*}

\subsection{Transfer to a target excluded from training}\label{sec:excluded}

The \(^{235}\mathrm{U}(n,f)\) calculations provide a separate transfer test because this target nucleus is excluded from training and model selection. Figures~\ref{fig:uthermal} and \ref{fig:u15} show representative predictions at thermal incident energy and at \(E_n=15\) MeV. All four models reproduce the main peak of the GEF distribution, with differences in peak height, width, and surrounding spectral shape.

The corresponding spectrum-level errors relative to GEF are summarized in Table~\ref{tab:physicalu235}. BMDN gives the smallest mean-TKE error, while GPR gives the smallest width error. For the nominal 95\% pointwise predictive intervals, coverage is 54.3\% for GPR and 94.9\% for GPR-WK, compared with \(98.4\pm0.7\%\) for MDN and \(99.1\pm0.5\%\) for BMDN. The higher neural-model coverage is accompanied by substantially wider predictive intervals.

\begin{table}[!t]
\caption{Spectrum-level mean absolute errors relative to GEF for the excluded \(^{235}\mathrm{U}(n,f)\) spectra. Mean-TKE and width errors are calculated using Eq.~(\ref{eq:spectrum_mae}). MDN and BMDN values are the mean \(\pm\) sample standard deviation over three independent fits; each GP model uses a single final fit. All values are in MeV.}
\label{tab:physicalu235}
\centering
\begingroup
\setlength{\tabcolsep}{12pt}
\renewcommand{\arraystretch}{1.12}
\begin{tabular}{lrr}
\toprule
Model & Mean-TKE MAE & Width MAE\\
\midrule
GPR & 0.558 & 0.308\\
GPR-WK & 0.621 & 0.371\\
MDN & $0.571\pm0.115$ & $0.660\pm0.238$\\
BMDN & $0.362\pm0.035$ & $0.750\pm0.125$\\
\bottomrule
\end{tabular}
\endgroup
\end{table}

\subsection{Experimental illustration and motivation for the extended analysis}
\begin{figure*}[!t]
\centering
\includegraphics[width=.97\textwidth]{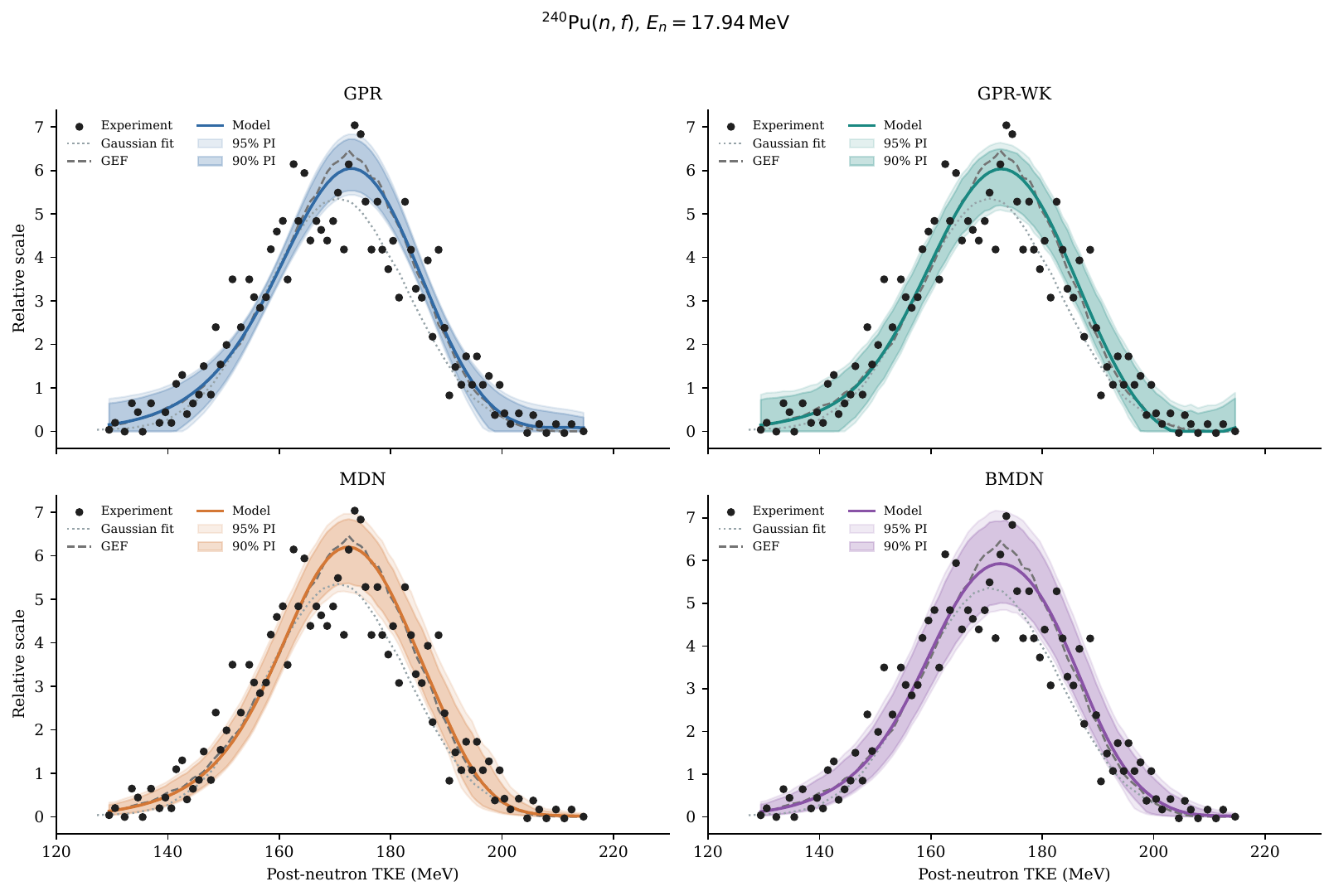}
\caption{Post-neutron TKE distribution for \(^{240}\mathrm{Pu}(n,f)\) at the experimental group centre \(E_n=17.94\) MeV, comparing the experimental spectrum with GEF and the four model predictions. The model curves are normalized positive point predictions. Dark and light shading denote the 90\% and 95\% pointwise predictive intervals of the normalized spectral shape defined in Eq.~(\ref{eq:shapePI}).}
\label{fig:pu17}
\end{figure*}

\begin{figure*}[!t]
\centering
\includegraphics[width=.97\textwidth]{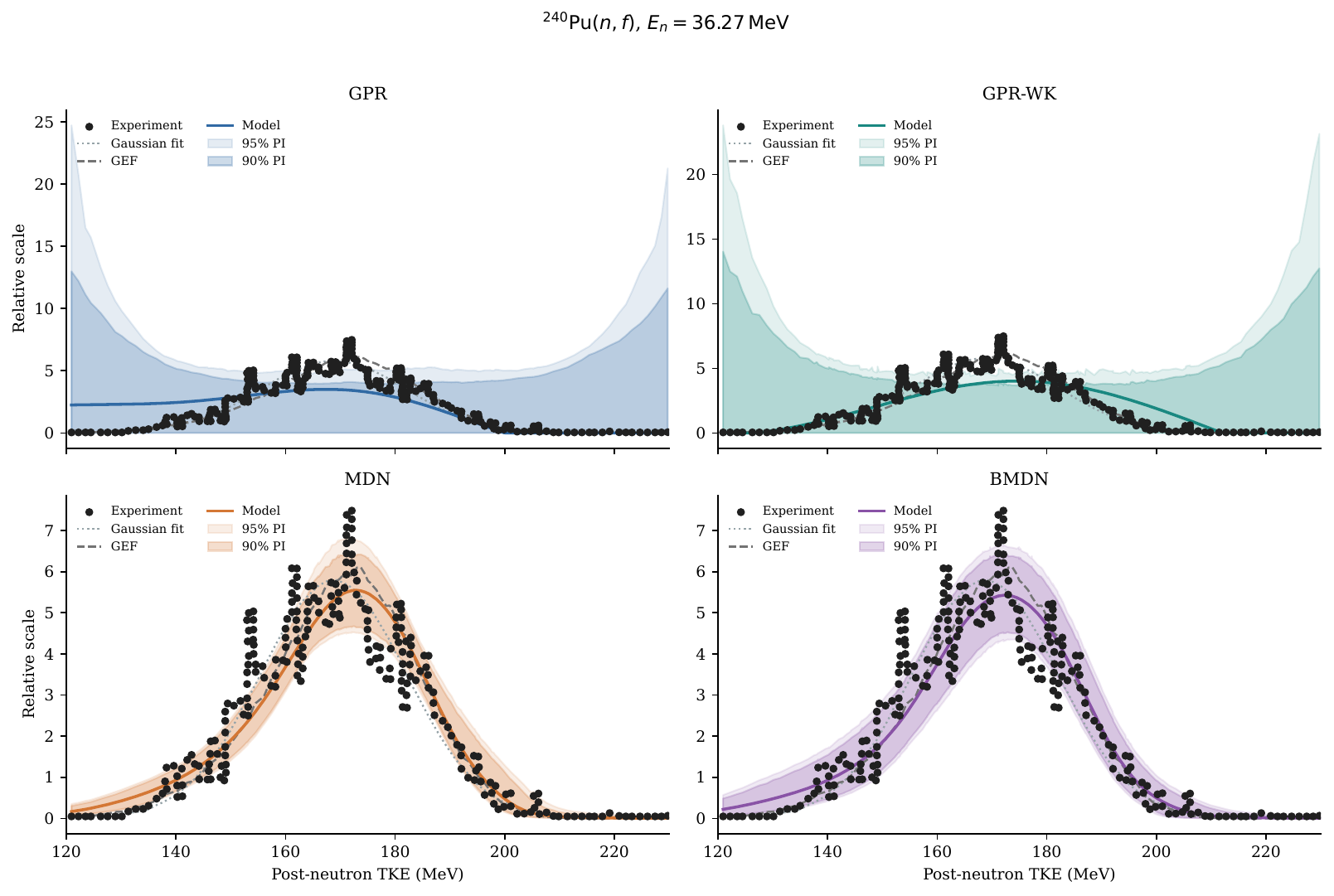}
\caption{Post-neutron TKE distribution for \(^{240}\mathrm{Pu}(n,f)\) at \(E_n=36.27\) MeV, above the 20-MeV training limit, comparing the experimental spectrum with GEF and the four model predictions. Shaded regions are 90\% and 95\% normalized-shape predictive intervals constructed from complete predictive spectra as in Eq.~(\ref{eq:shapePI}). For the GP models, only posterior realizations with positive area after the nonnegative projection contribute to the normalized-shape intervals.}
\label{fig:highfour36}
\end{figure*}

The \(^{240}\mathrm{Pu}(n,f)\) measurements provide an external illustration of the four-model predictions. This target is excluded from the four-model fitting population, and the two representative cases in Figs.~\ref{fig:pu17} and \ref{fig:highfour36} probe different energy regimes. The 17.94-MeV case lies within the incident-energy range represented in training, whereas 36.27 MeV lies above the 20-MeV training limit and therefore combines transfer in target identity with extrapolation in incident energy.

At 17.94 MeV, the four models reproduce the main experimental spectral structure with model-dependent differences in peak height, width, and uncertainty. At 36.27 MeV, the predictions become more sensitive to the model, particularly for the GP models.

The four-model study shows that the relative performance of the models depends on whether pointwise reconstruction, spectrum-level observables, or predictive-interval behavior is considered. It also  indicates that incident-energy extrapolation is more demanding than interpolation within the original training domain. The dependence on the available energy range is therefore examined systematically with MDN and BMDN in the extended-domain analysis below.

\clearpage

\section{Extended-domain analysis}\label{sec:full21}

The four-model comparison shows that predictions become more model dependent when the incident energy lies outside the original \(E_n\le20\) MeV training domain. The extended-domain analysis is therefore designed to separate two questions: whether MDN and BMDN continue to reproduce the underlying \gef\ calculations when the calculated domain is enlarged, and how the energy range represented in training affects comparison with measured high-energy TKE distributions.

\subsection{Extended-domain dataset and model setup}\label{sec:fulldata}

The extended-domain analysis uses 1,025 complete GEF TKE distributions for 20 target nuclei, corresponding to 107,191 sampled spectral values. The available incident-neutron energies are target dependent and extend to approximately 76 MeV. Most target nuclei are represented only up to 40 MeV; \(^{242}\mathrm{Pu}\) is the only training target with calculated spectra above this energy. An additional 53 \(^{235}\mathrm{U}(n,f)\) distributions are excluded from training and model selection and are used for evaluation. The target and energy coverage is summarized in Table~\ref{tab:fullcoverage}.

\begin{table*}[!t]

\caption{Target nuclei used for extended-domain training, with the number of complete GEF TKE distributions and the available incident-neutron-energy range. The \(^{235}\mathrm{U}(n,f)\) evaluation population is excluded from training. Only \(^{242}\mathrm{Pu}\) extends above 40 MeV; the listed ranges do not imply that every intermediate energy is available.}

\label{tab:fullcoverage}

\centering

\begingroup

\setlength{\tabcolsep}{10pt}

\renewcommand{\arraystretch}{1.06}

\begin{tabular}{lrr@{\qquad}lrr}
\toprule
Target & Spectra & $E_n$ (MeV) & Target & Spectra & $E_n$ (MeV) \\
\midrule
$^{227}\mathrm{Th}$ & 52 & $2.53\!\times\!10^{-8}$--40 &
$^{242}\mathrm{Pu}$ & 64 & $2.53\!\times\!10^{-8}$--76 \\
$^{229}\mathrm{Th}$ & 53 & $2.53\!\times\!10^{-8}$--40 &
$^{241}\mathrm{Am}$ & 51 & $2.53\!\times\!10^{-8}$--40 \\
$^{232}\mathrm{Th}$ & 53 & $2.53\!\times\!10^{-8}$--40 &
$^{243}\mathrm{Am}$ & 40 & 1--40 \\
$^{230}\mathrm{Pa}$ & 40 & 1--40 &
$^{244}\mathrm{Cm}$ & 51 & $2.53\!\times\!10^{-8}$--40 \\
$^{231}\mathrm{Pa}$ & 53 & $2.53\!\times\!10^{-8}$--40 &
$^{245}\mathrm{Cm}$ & 51 & $2.53\!\times\!10^{-8}$--40 \\
$^{233}\mathrm{U}$ & 53 & $2.53\!\times\!10^{-8}$--40 &
$^{248}\mathrm{Cm}$ & 53 & $2.53\!\times\!10^{-8}$--40 \\
$^{234}\mathrm{U}$ & 53 & $2.53\!\times\!10^{-8}$--40 &
$^{251}\mathrm{Cf}$ & 53 & $2.53\!\times\!10^{-8}$--40 \\
$^{238}\mathrm{U}$ & 53 & $2.53\!\times\!10^{-8}$--40 &
$^{253}\mathrm{Es}$ & 40 & 1--40 \\
$^{237}\mathrm{Np}$ & 53 & $2.53\!\times\!10^{-8}$--40 &
$^{254}\mathrm{Es}$ & 53 & $2.53\!\times\!10^{-8}$--40 \\
$^{239}\mathrm{Pu}$ & 53 & $2.53\!\times\!10^{-8}$--40 &
$^{255}\mathrm{Fm}$ & 53 & $2.53\!\times\!10^{-8}$--40 \\
\bottomrule
\end{tabular}

\endgroup

\end{table*}

The dataset is divided into three independent spectrum-grouped partitions, with complete \((Z,A,E_n)\) distributions assigned to mutually exclusive fitting, validation, and test subsets. Within each partition, nested training populations containing approximately 25\%, 50\%, 75\%, and 100\% of the available fitting spectra are considered, while the validation and test subsets remain fixed. Each neural model is trained with three independent initializations, giving nine training configurations for each prescribed training population.

The MDN and BMDN retain the same input variables and mixture representation used in the four-model comparison. Training uses Adam with an initial learning rate of \(3\times10^{-3}\), a batch size of 256, and early stopping based on validation loss. For BMDN, the KL contribution is scaled as \(1/N_{\rm fit}\), where \(N_{\rm fit}\) is the number of fitting observations. Exact GPR is not used for the extended dataset because its covariance-matrix factorization becomes computationally prohibitive at this scale. 

\subsection{Extended-domain performance}\label{sec:fullperformance}

\begin{table*}[t]
\caption{Pointwise count-reconstruction and 95\% predictive-interval diagnostics from the extended-domain analysis, averaged over nine training configurations at each training fraction. The reported \(\pm\) values are sample standard deviations of the RMSE across configurations and do not represent experimental uncertainties. Smaller interval scores indicate a better balance between predictive-interval width and deviations of GEF values outside the interval.}
\label{tab:fullscores}
\centering
\begingroup
\setlength{\tabcolsep}{7pt}
\renewcommand{\arraystretch}{1.12}
\begin{tabular}{llrrrr}
\toprule
Population & Fraction & Model & RMSE & Coverage (\%) & Interval score\\
\midrule
Test set & 25\% & MDN  & $141.75\pm16.05$ & 94.61 & 482.06\\
     &       & BMDN & $177.06\pm23.94$ & 96.68 & 674.12\\
Test set & 50\% & MDN  & $130.17\pm9.17$  & 94.74 & 411.22\\
     &       & BMDN & $143.08\pm21.07$ & 96.16 & 511.61\\
Test set & 75\% & MDN  & $124.56\pm14.47$ & 94.91 & 399.23\\
     &       & BMDN & $136.09\pm13.67$ & 95.90 & 434.49\\
Test set & 100\% & MDN  & $123.37\pm13.89$ & 94.85 & 407.69\\
     &        & BMDN & $133.97\pm9.96$  & 95.80 & 419.61\\
\midrule
$^{235}\mathrm{U}(n,f)$ & 25\% & MDN  & $133.21\pm6.72$  & 96.81 & 508.09\\
                         &       & BMDN & $128.72\pm5.94$  & 97.80 & 751.60\\
$^{235}\mathrm{U}(n,f)$ & 50\% & MDN  & $130.58\pm7.01$  & 94.84 & 447.45\\
                         &       & BMDN & $116.68\pm8.79$  & 97.49 & 519.93\\
$^{235}\mathrm{U}(n,f)$ & 75\% & MDN  & $131.13\pm12.49$ & 94.30 & 447.52\\
                         &       & BMDN & $117.72\pm5.23$  & 95.31 & 474.11\\
$^{235}\mathrm{U}(n,f)$ & 100\% & MDN  & $126.46\pm7.72$  & 94.56 & 462.45\\
                         &        & BMDN & $113.98\pm2.71$  & 95.62 & 447.15\\
\bottomrule
\end{tabular}
\endgroup
\end{table*}

Increasing the amount of calculated training data generally improves reconstruction of the GEF spectra, although the dependence is not strictly monotonic for every model and dataset. On the independent test set, the MDN RMSE decreases from \(141.75\pm16.05\) counts at the 25\% training fraction to \(123.37\pm13.89\) counts at the full fraction, while the corresponding BMDN RMSE decreases from \(177.06\pm23.94\) to \(133.97\pm9.96\) counts. For the excluded \(^{235}\mathrm{U}(n,f)\) spectra, BMDN decreases from \(128.72\pm5.94\) to \(113.98\pm2.71\) counts, whereas MDN shows a smaller net reduction from \(133.21\pm6.72\) to \(126.46\pm7.72\) counts. The complete training-fraction dependence is summarized in Table~\ref{tab:fullscores}.

Predictive-interval coverage provides complementary information to the mean count errors. As shown in Table~\ref{tab:fullscores}, the nominal 95\% coverage remains close to 95\% over most of the training-fraction sequence. On the independent test set, coverage ranges from 94.61--94.91\% for MDN and from 95.80--96.68\% for BMDN. For the excluded \(^{235}\mathrm{U}(n,f)\) spectra, the initially higher BMDN coverage decreases from 97.80\% at the 25\% fraction to 95.62\% at the full fraction. The interval scores generally decrease as the training population is enlarged, particularly for BMDN, indicating improved sharpness while maintaining coverage near the nominal level.

For BMDN, the conditional-mixture term remains the dominant contribution to the modeled predictive variance throughout the extended-domain analysis. On the independent test set, its average contribution increases from 96.28\% at the 25\% training fraction to 97.57\% at the full fraction, while the variational-weight contribution decreases from 3.72\% to 2.43\%. For the excluded \(^{235}\mathrm{U}(n,f)\) spectra, the corresponding contributions change from 97.14\% to 98.11\% and from 2.86\% to 1.89\%, respectively.

Spectrum-level errors relative to GEF are calculated with Eq.~(\ref{eq:spectrum_mae}) for each of the nine training configurations and then averaged across configurations. The resulting mean-TKE and width errors are reported in Table~\ref{tab:fullphysics}. On the test set, MDN gives a slightly smaller mean-TKE error, while the two models give nearly identical width errors. For the excluded \(^{235}\mathrm{U}(n,f)\) spectra, MDN gives the slightly smaller mean-TKE error and BMDN the smaller width error.

\begin{table}[!t]

\caption{Mean absolute errors in the mean TKE and distribution width relative to GEF at the full training fraction. Values are averages over nine training configurations, with sample standard deviations across configurations. All values are in MeV.}

\label{tab:fullphysics}

\centering

\begingroup

\setlength{\tabcolsep}{5pt}

\renewcommand{\arraystretch}{1.10}

\begin{tabular}{llrr}
\toprule
Population & Model & Mean-TKE MAE & Width MAE\\
\midrule
Test set & MDN  & $0.730\pm0.074$ & $0.384\pm0.057$\\
     & BMDN & $0.821\pm0.063$ & $0.385\pm0.036$\\
$^{235}\mathrm{U}$ & MDN  & $0.732\pm0.043$ & $0.540\pm0.097$\\
                   & BMDN & $0.761\pm0.053$ & $0.481\pm0.052$\\
\bottomrule
\end{tabular}

\endgroup

\end{table}

\begin{figure*}[t]

\centering

\includegraphics[width=.89\textwidth]{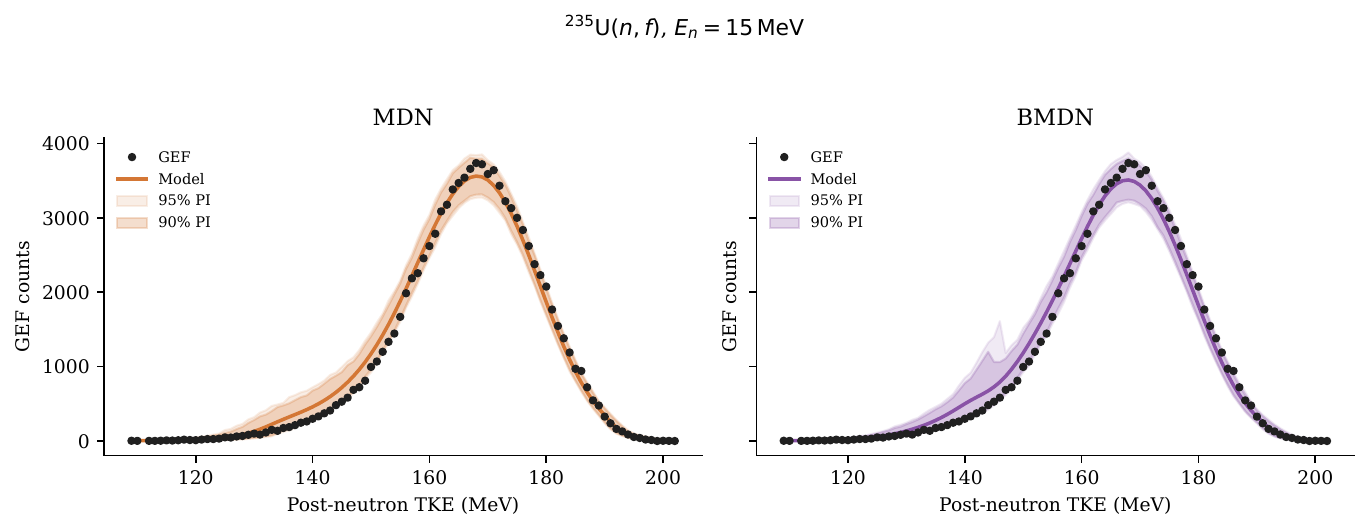}

\caption{Post-neutron TKE distribution for \(^{235}\mathrm{U}(n,f)\) at \(E_n=15\) MeV from the extended-domain analysis. Filled black markers show GEF spectral values, solid orange and purple curves show the MDN and BMDN predictions, and shaded regions denote the corresponding 90\% and 95\% predictive intervals on the GEF count scale.}

\label{fig:fullu235}

\end{figure*}

Figure~\ref{fig:fullu235} shows a representative extended-domain prediction of the TKE distribution for the excluded \(^{235}\mathrm{U}(n,f)\) target at \(E_n=15\) MeV. Both MDN and BMDN reproduce the main structure of the calculated distribution, with model-dependent differences around the peak and in the predictive-interval widths. The full excluded-target evaluation includes 53 \(^{235}\mathrm{U}(n,f)\) TKE distributions spanning the available incident-energy range. 

\subsection{Comparison with experiment}\label{sec:full60five}

The full-range MDN and BMDN models are compared with the 60 measured post-neutron TKE distributions listed in Table~\ref{tab:exp60coverage}. For each target \(t\), Eq.~(\ref{eq:spectrum_mae}) is evaluated on the corresponding experimental subset \(\mathcal S_t\), using the experimental mean TKE or width as the reference. Here \(N_t=20\) for \(^{235}\)U and \(N_t=10\) for each of the other four targets. Neural-model values are calculated separately for each of the nine training configurations and then averaged across configurations, whereas GEF provides a single value for each experimental case.

The errors are summarized in Table~\ref{tab:full60targets}. The model with the smallest numerical error changes with both target and observable. 

\begin{table*}[!t]

\caption{Mean absolute errors relative to experiment for GEF and the full-range MDN/BMDN predictions. Experimental widths are obtained from the reported TKE variances as \(\sigma_T=\sqrt{\sigma_T^2}\); GEF and model widths are calculated from the corresponding normalized spectra using Eq.~(\ref{eq:meanwidth}). Neural entries are averages over nine training configurations. All values are in MeV.}

\label{tab:full60targets}

\centering

\begingroup

\setlength{\tabcolsep}{10pt}

\renewcommand{\arraystretch}{1.08}

\begin{tabular}{lrrrrrr}
\toprule
& \multicolumn{3}{c}{Mean TKE} & \multicolumn{3}{c}{Width}\\
\cmidrule(lr){2-4}\cmidrule(l){5-7}
Target & \gef & MDN & BMDN & \gef & MDN & BMDN\\
\midrule
$^{235}$U  & 0.781 & 0.696 & 0.553 & 1.086 & 1.446 & 1.155\\
$^{237}$Np & 0.956 & 0.994 & 0.914 & 2.084 & 2.193 & 1.878\\
$^{239}$Pu & 0.824 & 0.810 & 0.948 & 2.277 & 2.101 & 1.905\\
$^{240}$Pu & 0.712 & 0.968 & 1.124 & 1.366 & 1.213 & 1.184\\
$^{242}$Pu & 1.972 & 2.064 & 2.107 & 1.386 & 1.298 & 1.308\\
\bottomrule
\end{tabular}

\endgroup

\end{table*}

\begin{figure*}[t]

\centering

\includegraphics[width=.98\textwidth]{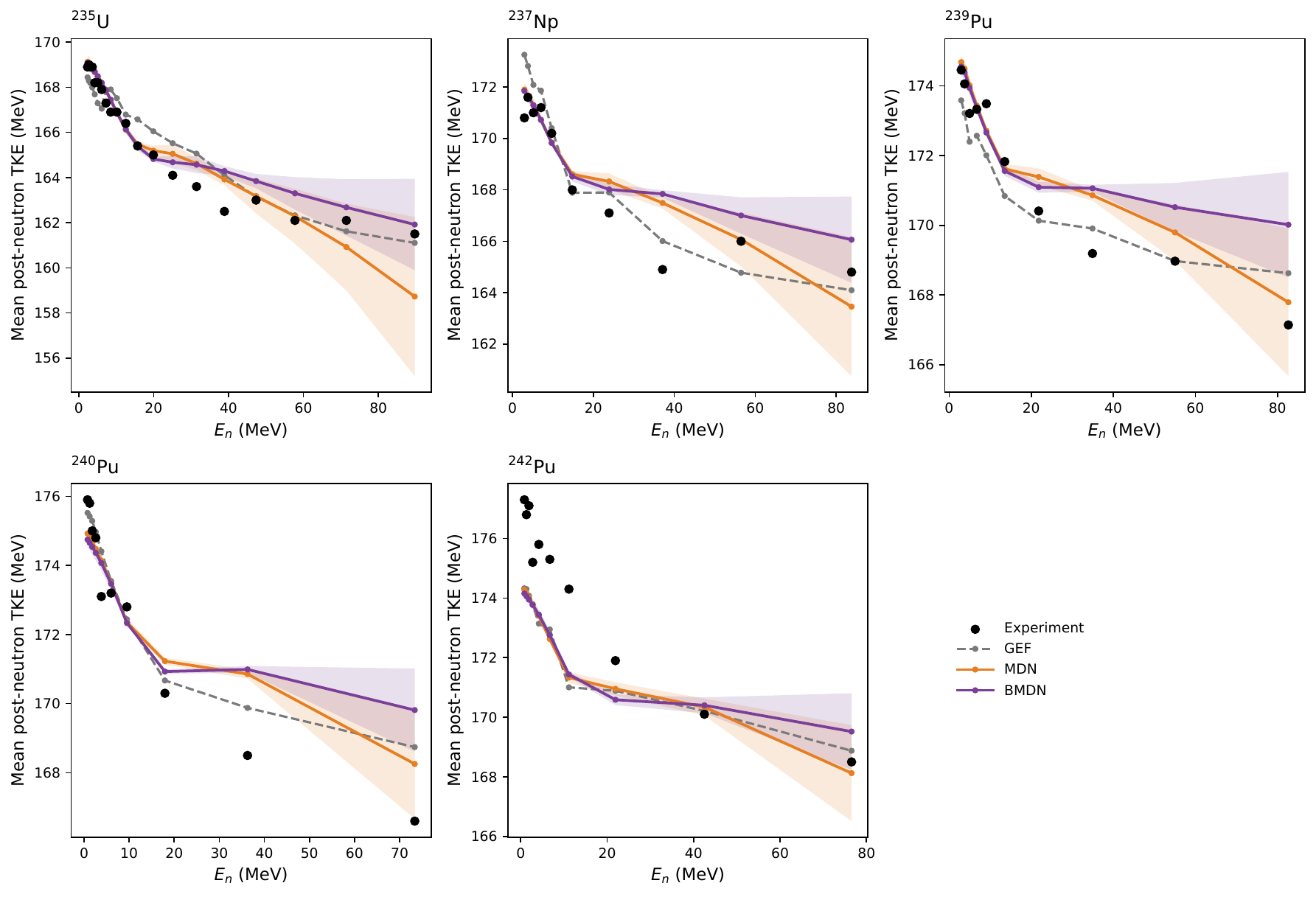}

\caption{Experimental mean post-neutron TKE for the five experimental targets compared with GEF and the full-range MDN/BMDN predictions. Experimental values are shown as black symbols, GEF as a gray dashed curve with markers, MDN in orange, and BMDN in purple. Neural curves show means over nine training configurations, and the shading indicates the corresponding sample standard deviation.}

\label{fig:all5mean}

\end{figure*}

\begin{figure*}[t]

\centering

\includegraphics[width=.98\textwidth]{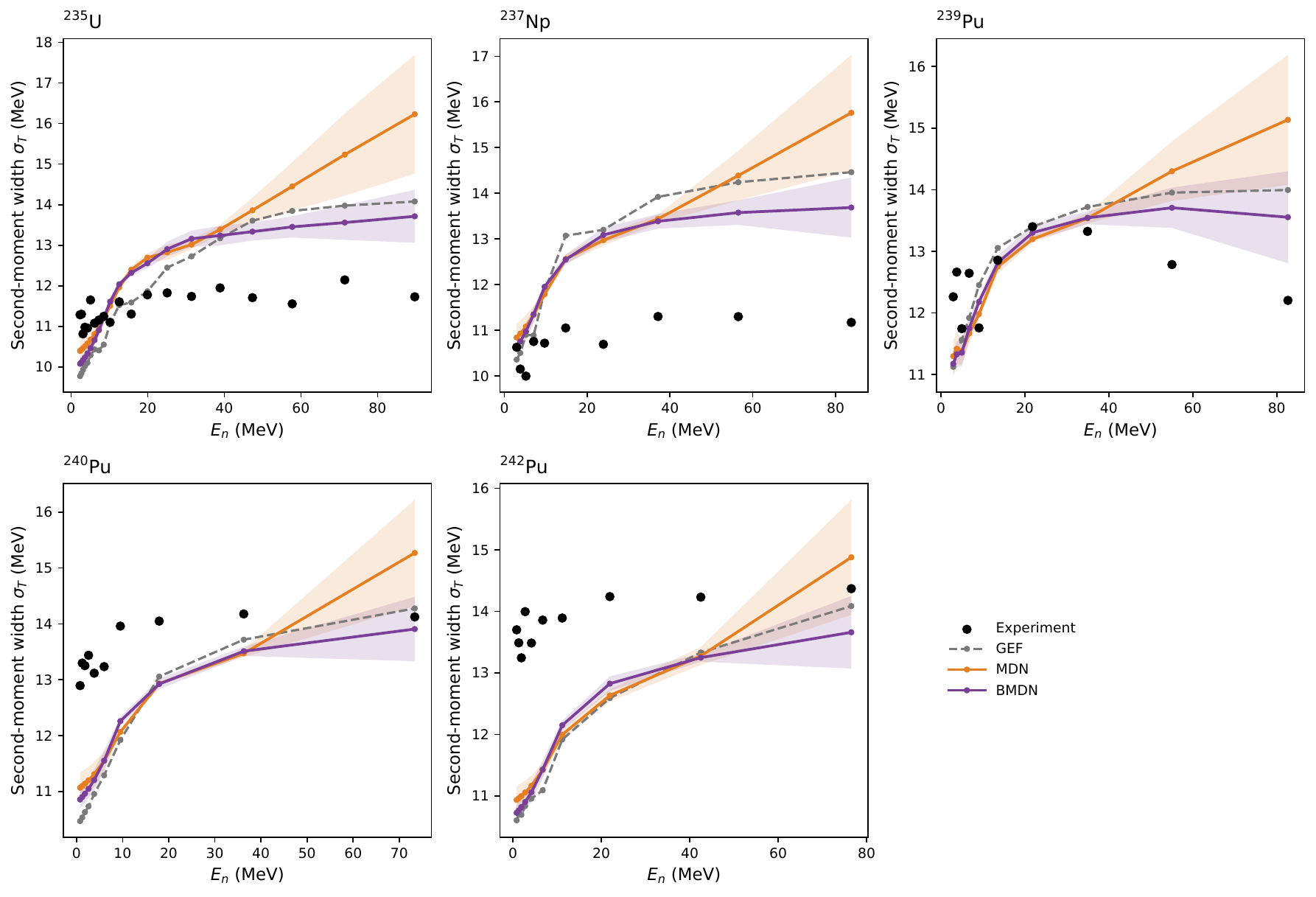}

\caption{Second-moment widths for the five experimental targets, evaluated on the experimental TKE coordinates for experiment, GEF, and the full-range MDN/BMDN predictions. Neural curves show means over nine training configurations, and the shading indicates the corresponding sample standard deviation. This common-grid comparison is used as a sensitivity test to the width definition.}

\label{fig:all5width}

\end{figure*}

Figures~\ref{fig:all5mean} and \ref{fig:all5width} show the energy dependence of the mean TKE and distribution width for the five experimental targets. The mean-TKE comparison uses the values reported in the experimental studies. For the width sensitivity test, experiment, GEF, and model predictions are evaluated using the same second-moment definition on the experimental TKE coordinates. The neural curves show averages over the nine full-range training configurations, while the shaded regions represent their sample standard deviation.

\begin{figure*}[!t]
\centering
\includegraphics[width=.95\textwidth]{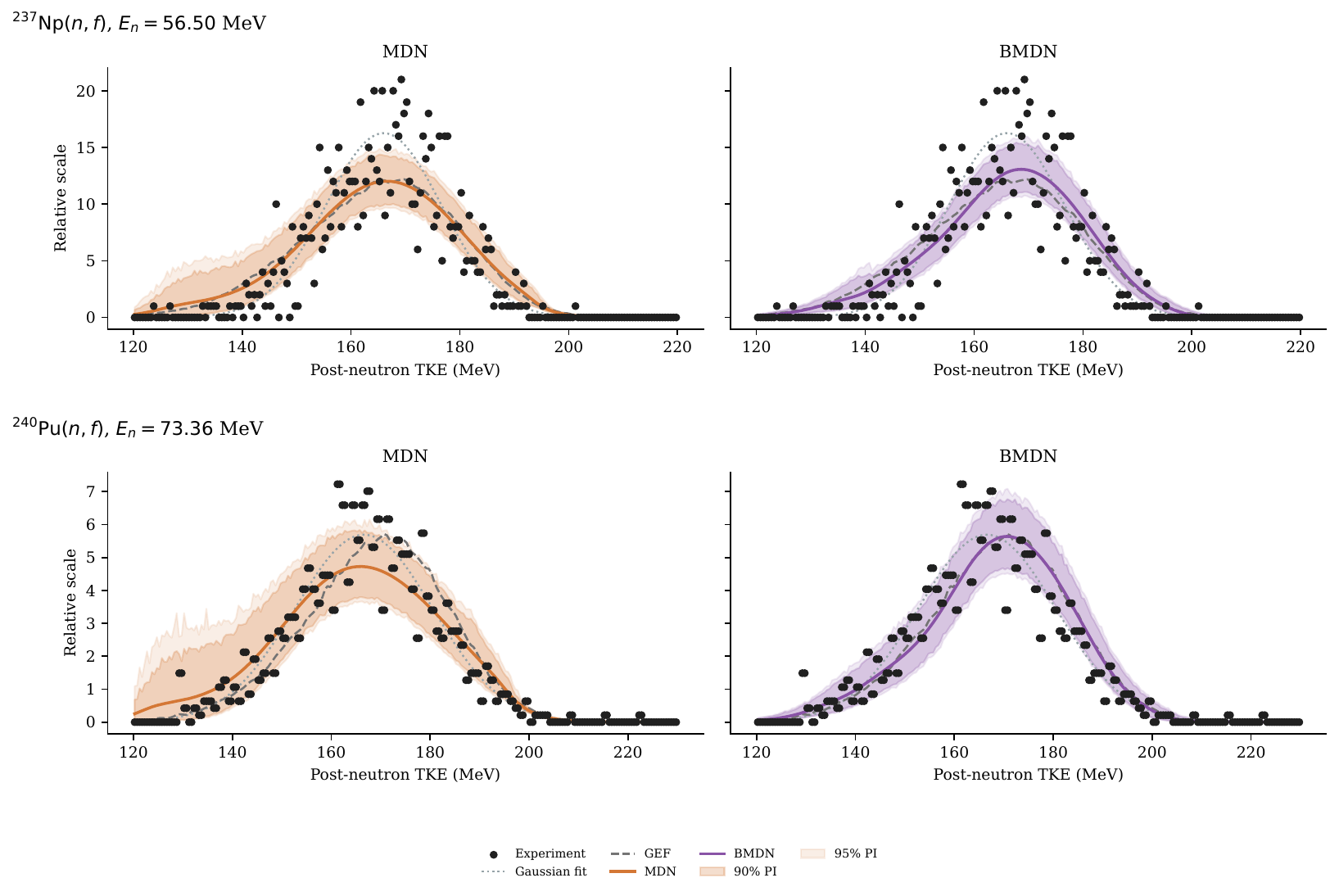}
\caption{Post-neutron TKE distributions. The upper row shows \(^{237}\mathrm{Np}(n,f)\) at \(E_n=56.50\) MeV and the lower row \(^{240}\mathrm{Pu}(n,f)\) at \(E_n=73.36\) MeV; left and right panels show MDN and BMDN, respectively. Black symbols denote the experimental distributions, the gray dotted curves the Gaussian fits, and the gray dashed curves the direct \gef\ calculations. The orange and purple curves show the MDN and BMDN predictions, respectively, while dark and light shading denote the nominal 90\% and 95\% pointwise predictive intervals of the normalized spectral shape, constructed according to Eq.~(\ref{eq:shapePI}). The intervals are shown for one predeclared representative full-training configuration and quantify within-fit predictive uncertainty rather than variability across independent training configurations.}
\label{fig:full21spectra}
\end{figure*}

Figure~\ref{fig:full21spectra} shows representative high-energy predictions from the extended-domain analysis for two distinct generalization settings. For \(^{237}\mathrm{Np}(n,f)\) at \(E_n=56.50\) MeV, the target nucleus is represented in training, but the experimental energy lies above its target-specific training range, which extends to 40 MeV. For \(^{240}\mathrm{Pu}(n,f)\) at \(E_n=73.36\) MeV, the target nucleus is excluded from fitting and the incident energy lies near the upper edge of the overall training-energy domain. The shaded regions show within-fit normalized-shape predictive intervals from one predeclared representative full-training configuration; the numerical conclusions are based on the complete set of nine training configurations. These intervals are distinct from the across-configuration sample standard deviations shown in Figs.~\ref{fig:all5mean} and \ref{fig:all5width}.

At high incident energy, the distribution width, centroid, and complete spectral shape do not necessarily show the same accuracy ordering. Figure~\ref{fig:full21spectra} illustrates the complete spectral profiles for the \(^{237}\mathrm{Np}\) and \(^{240}\mathrm{Pu}\) cases, while Table~\ref{tab:highfull} gives the corresponding spectrum-level comparison for \(^{240}\mathrm{Pu}\) at 36.27 and 73.36 MeV. At 73.36 MeV, BMDN gives a width of \(13.91\pm0.57\) MeV, close to the experimental value of 13.92 MeV, whereas the ordering of the mean-TKE errors differs and the spectral comparison shows additional differences in shape. The centroid, width, and complete distribution are therefore assessed separately.

\begin{table*}[!t]

\caption{High-energy \(^{240}\mathrm{Pu}(n,f)\) comparison at 36.27 and 73.36 MeV. MDN and BMDN values are means \(\pm\) sample standard deviations across the nine full-range training configurations. All values are in MeV.}

\label{tab:highfull}

\centering

\begingroup

\setlength{\tabcolsep}{10pt}

\renewcommand{\arraystretch}{1.08}

\begin{tabular}{lrrrr}
\toprule
& \multicolumn{2}{c}{36.27 MeV} & \multicolumn{2}{c}{73.36 MeV}\\
\cmidrule(lr){2-3}\cmidrule(l){4-5}
Source & $\langle T\rangle$ & $\sigma_T$ & $\langle T\rangle$ & $\sigma_T$\\
\midrule
Experimental data & 168.50 & 14.15 & 166.60 & 13.92\\
\gef              & 169.88 & 13.77 & 168.74 & 14.34\\
MDN               & $170.86\pm0.13$ & $13.46\pm0.07$ &
                    $168.25\pm1.59$ & $15.27\pm0.95$\\
BMDN              & $170.99\pm0.10$ & $13.50\pm0.08$ &
                    $169.81\pm1.20$ & $13.91\pm0.57$\\
\bottomrule
\end{tabular}

\endgroup

\end{table*}

To distinguish improvement relative to experiment from recovery of the underlying calculated response, the same high-energy widths were also compared directly with GEF. For \(E_n>40\) MeV, the BMDN width MAE relative to GEF decreases from 3.48 MeV for \(E_n\le20\) MeV training to 0.83 MeV for \(E_n\le40\) MeV and 0.47 MeV for the full-range model. Thus, extending the training domain systematically improves reproduction of the calculated GEF width even though the corresponding agreement with experiment is not monotonic.

\subsection{Dependence on the training-energy range}\label{sec:coverage60}

The strongest dependence on the calculated training range is observed for the high-energy TKE-distribution width. For a threshold \(E_c\), define the experimental subset
\[
\mathcal H(E_c)=\{i:E_{n,i}>E_c\}.
\]
The width MAE on this common subset is Eq.~(\ref{eq:spectrum_mae}) with \(q=\sigma_T\) and \(\mathcal S=\mathcal H(E_c)\); the neural values reported below are averaged over the nine training configurations. For \(E_c=40\) MeV, \(N_{E_c}=11\). The BMDN width MAE decreases from 5.63 MeV when training is restricted to \(E_n\le20\) MeV to 1.76 MeV for \(E_n\le40\) MeV and 1.90 MeV for the full calculated range. The corresponding MDN values are 7.04, 3.09, and 3.02 MeV. The GEF width MAE for the same 11 spectra is 2.32 MeV.

Matched-size controls show that the improvement is not explained simply by the larger number of calculated spectra. When the number of complete fitting and validation spectra is held equal to that of the \(E_n\le20\) MeV case, extending the BMDN training range through 40 MeV gives a high-energy width MAE of 1.77 MeV, essentially identical to the 1.76 MeV obtained when all available spectra through 40 MeV are used.

The matched-size \(E_n\le30\) MeV controls give intermediate errors. Above 40 MeV, the two independently predetermined 30-MeV realizations give BMDN width MAEs of 2.70 and 2.53 MeV, compared with 5.63 MeV for \(E_n\le20\) MeV and 1.77 MeV for the matched \(E_n\le40\) MeV case. The intermediate 30-MeV controls therefore show that the reduction in width error is already evident before the training range reaches 40 MeV. The complete training-range comparison is summarized in Table~\ref{tab:matched30threshold}.

\begin{table*}[!t]

\caption{Training-range study for BMDN and MDN. Matched-size cases preserve, for each target, the number of complete fitting and validation spectra used in the \(E_n\le20\) MeV training population while extending the available training energies to 30 or 40 MeV. Entries are configuration-averaged width MAEs evaluated with Eq.~(\ref{eq:spectrum_mae}); the \(E_n>30\) and \(E_n>40\) columns contain 16 and 11 experimental spectra, respectively. All values are in MeV.}

\label{tab:matched30threshold}

\centering

\begingroup

\setlength{\tabcolsep}{8pt}

\renewcommand{\arraystretch}{1.08}

\begin{tabular}{llrr}
\toprule
Model & Training range & $E_n>30$ & $E_n>40$\\
\midrule
BMDN & $E_n\le20$ MeV & 4.961 & 5.629\\
BMDN & $E_n\le30$ MeV (matched-size A) & 2.515 & 2.695\\
BMDN & $E_n\le30$ MeV (matched-size B) & 2.325 & 2.527\\
BMDN & $E_n\le40$ MeV (matched-size) & 1.792 & 1.769\\
BMDN & $E_n\le40$ MeV (all spectra) & 1.773 & 1.762\\
BMDN & Full range ($\sim76$ MeV) & 1.878 & 1.903\\
\midrule
MDN & $E_n\le20$ MeV & 5.945 & 7.043\\
MDN & $E_n\le30$ MeV (matched-size A) & 4.338 & 5.293\\
MDN & $E_n\le30$ MeV (matched-size B) & 3.922 & 4.781\\
MDN & $E_n\le40$ MeV (matched-size) & 2.898 & 3.373\\
MDN & $E_n\le40$ MeV (all spectra) & 2.704 & 3.092\\
MDN & Full range ($\sim76$ MeV) & 2.651 & 3.018\\
\bottomrule
\end{tabular}

\endgroup

\end{table*}

\begin{figure*}[t]

\centering

\includegraphics[width=.82\textwidth]{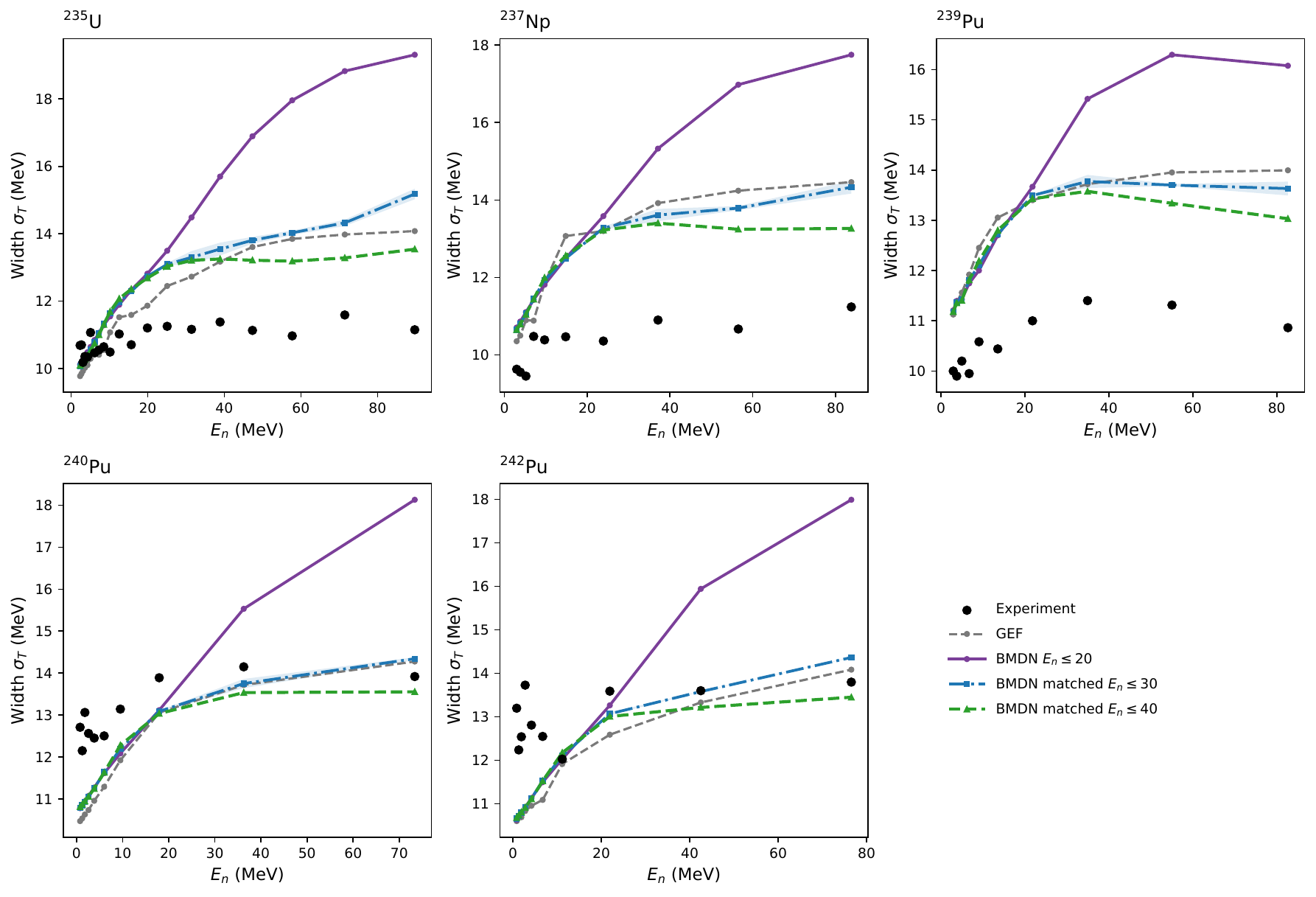}

\caption{Experimental widths, GEF widths, and BMDN predictions for the five experimental targets under different training-energy ranges. The \(E_n\le20\), matched \(E_n\le30\), and matched \(E_n\le40\) MeV cases preserve the same numbers of complete fitting and validation spectra per target. For the matched \(E_n\le30\) MeV case, the curve shows the mean of the two independently predetermined matched-size realizations, while the shaded region spans their two configuration-averaged predictions at each incident energy. The intermediate 30-MeV result lies between the 20- and 40-MeV cases, showing that the improvement is already present before the training range reaches 40 MeV.}

\label{fig:matchedmechanism}

\end{figure*}

Figure~\ref{fig:matchedmechanism} shows the corresponding energy dependence. The BMDN trained only through 20 MeV develops increasingly broad high-energy distributions for several targets. For the matched \(E_n\le30\) MeV case, the curve represents the mean of the two independently predetermined realizations listed separately in Table~\ref{tab:matched30threshold}, while the shaded region spans their configuration-averaged predictions at each incident energy. The matched-30 result lies between the \(E_n\le20\) and matched \(E_n\le40\) MeV cases over the high-energy region, while the matched \(E_n\le40\) MeV model follows the GEF width trend substantially more closely despite using the same numbers of complete fitting and validation spectra as the \(E_n\le20\) MeV model.

\begin{table*}[!t]

\caption{Differences in the high-energy BMDN width MAE between selected training-energy ranges for the \(E_n>40\) MeV experimental subset. Positive values indicate a lower width MAE for the second training condition. The corresponding 95\% intervals are shown for both two-way and target-cluster resampling. All values are in MeV.}

\label{tab:mainrobustness}

\centering

\begingroup

\small

\setlength{\tabcolsep}{5pt}

\renewcommand{\arraystretch}{1.08}

\begin{tabular}{p{0.39\textwidth}rrr}
\toprule
Contrast & Difference & 95\% two-way interval & 95\% target-cluster interval\\
\midrule
BMDN \(E_n\le20 - E_n\le40\) (all spectra), width, \(E_n>40\) & 3.867 & [2.850, 4.967] & [2.57, 5.01]\\
BMDN \(E_n\le20 - E_n\le40\) matched-size, width, \(E_n>40\) & 3.859 & [2.787, 4.996] & [2.496, 5.039]\\
BMDN \(E_n\le40 -\) full, width, \(E_n>40\) & -0.141 & [-0.459, 0.143] & [-0.48, 0.18]\\
\bottomrule
\end{tabular}

\endgroup

\end{table*}

The differences in high-energy width MAE between the selected training conditions are summarized in Table~\ref{tab:mainrobustness}. The 95\% intervals are obtained using paired two-way resampling over training configurations and experimental spectra, together with a complementary target-cluster resampling over the five nuclei. For both the unrestricted and matched-size \(E_n\le20\) to \(E_n\le40\) MeV comparisons, the intervals remain positive, supporting a substantial reduction in width error when the training range is extended to 40 MeV. By contrast, the difference between the \(E_n\le40\) MeV and full-range models is consistent with zero.

This conclusion is retained in leave-one-out checks: the BMDN \(E_n\le20\)-to-\(E_n\le40\) reduction in high-energy width MAE remains 3.27--4.14 MeV when individual target nuclei are omitted, 3.77--4.01 MeV when individual grouped partitions are omitted, and 3.14--4.35 MeV when individual neural initializations are omitted.

The reduction in high-energy width error is also retained when the trainable BMDN prior is replaced by a fixed standard-normal prior. Above 40 MeV, the corresponding width MAEs are 5.41, 1.96, and 1.97 MeV for the \(E_n\le20\), \(E_n\le40\), and full-range training populations, respectively. The paired \(E_n\le20\)-to-\(E_n\le40\) reduction is 3.45 MeV, with a 95\% interval of 2.30--4.75 MeV, while the \(E_n\le40\)-to-full difference is consistent with zero.

The reduction is likewise retained when experiment, GEF, and model predictions are compared using a common second-moment width definition. Table~\ref{tab:widthdefinitions} shows the two width definitions side by side. Across the 60 experimental spectra, the experimental second-moment widths exceed the reported width parameters by 0.83 MeV on average (mean absolute difference 0.83 MeV); for the 11 spectra above 40 MeV the corresponding values are 0.64 and 0.66 MeV. The absolute MAEs therefore depend on the adopted width definition, but the large reduction between the 20- and 40-MeV training ranges is unchanged.

\begin{table*}[t]

\caption{High-energy (\(E_n>40\) MeV) width MAEs under the two width definitions used in the analysis. In the primary comparison, the experimental width is obtained from the reported TKE variance as \(\sigma_T=\sqrt{\sigma_T^2}\). The common-second-moment sensitivity test evaluates experiment, GEF, and model distributions using the same second-moment definition on the experimental TKE coordinates. All values are in MeV.}

\label{tab:widthdefinitions}

\centering

\begingroup

\setlength{\tabcolsep}{10pt}

\renewcommand{\arraystretch}{1.08}

\begin{tabular}{lrrrr}
\toprule
Width definition & BMDN \(\le20\) & BMDN \(\le40\) & BMDN full & GEF\\
\midrule
Experimental width from reported variance & 5.63 & 1.76 & 1.90 & 2.32\\
Common second moment & 4.99 & 1.39 & 1.47 & 1.72\\
\bottomrule
\end{tabular}

\endgroup

\end{table*}

The pointwise predictive intervals show a related dependence on training-energy coverage. Above 40 MeV, the BMDN coverage of GEF spectral values increases from 52.3\% for \(E_n\le20\) MeV to 74.7\% for \(E_n\le40\) MeV and 92.3\% for the full range. The corresponding MDN values are 23.0\%, 75.3\%, and 86.8\%. These coverage values are evaluated on the GEF spectral ordinates for the 11 \(E_n>40\) MeV target--energy cases entering the high-energy experimental comparison. Thus both reconstruction accuracy and predictive-interval coverage deteriorate strongly when the model is evaluated far beyond the training-energy range.

The high-energy GEF widths provide additional context for the training-range dependence. For each target, a simple carry-forward diagnostic was constructed by retaining the GEF width at the highest available \(E_n\le20\) MeV energy. At the 11 measured energies above 40 MeV, the actual GEF widths differ from these carry-forward values by 1.11--2.24 MeV, with a mean absolute difference of 1.66 MeV. In contrast, linear extrapolation of the low-energy GEF trend differs from the high-energy GEF widths by 6.88 MeV on average. Thus, although the GEF width is not constant, its high-energy evolution is comparatively slow.

The same carry-forward values give a width MAE of 1.41 MeV relative to experiment, compared with 2.32 MeV for GEF and 9.16 MeV for the linear extrapolation. The paired reduction in absolute error relative to GEF is 0.91 MeV, with a 95\% interval of 0.08--1.63 MeV; when the resampling is performed by target nucleus, the corresponding interval is \([-0.73,1.77]\) MeV. These results indicate that the improvement obtained as the training range is extended toward 40 MeV occurs against a comparatively slowly varying high-energy GEF width.

The mean TKE shows a different dependence on training range. Above 40 MeV, the BMDN mean-TKE MAE decreases from 3.07 to 2.34 to 1.43 MeV as the training range is extended from 20 to 40 MeV and then to the full range, while MDN changes from 4.57 to 1.40 to 1.44 MeV. The centroid results therefore do not follow the same pattern as the distribution widths, reinforcing the need to assess the two observables separately.

Finally, variation across the nine training configurations increases at high energy but remains smaller than the extrapolation error of the models trained only through 20 MeV. This spread characterizes the reproducibility of the neural predictions across training configurations.

\section{Discussion}\label{sec:discussion}

\subsection{Training-energy range and high-energy prediction}

The clearest result of the extended analysis is the strong dependence of the high-energy TKE-distribution width on the incident-neutron-energy range represented in training. Restricting the calculated training population to \(E_n\le20\) MeV leads to large width errors above 40 MeV, whereas extending the training range through 40 MeV substantially improves both MDN and BMDN predictions. The matched-size controls show that this improvement is not explained simply by the larger number of calculated spectra: for BMDN, the high-energy width MAE is 1.77 MeV for the matched \(E_n\le40\) MeV population and 1.76 MeV when all available spectra through 40 MeV are used.

The two matched \(E_n\le30\) MeV controls give intermediate errors between the 20- and 40-MeV cases. The intermediate 30-MeV controls therefore show that the reduction in width error is already evident before the training range reaches 40 MeV. The complete training-range comparison is summarized in Table~\ref{tab:matched30threshold}. The 20–40 MeV interval also overlaps the increasing importance of multichance fission in these systems. The present models do not include explicit fission-chance information, so the observed dependence reflects the changing TKE response encoded in the GEF calculations rather than an identified contribution from an individual fission channel.

Extending the training population beyond 40 MeV addresses a different regime. The learned models continue to approach the GEF response, but this does not lead to a corresponding systematic improvement relative to experiment. Once the model--GEF discrepancy becomes small, the remaining disagreement is increasingly determined by the difference between GEF and the measured distributions.

The GEF-based width diagnostics help place this behavior in context. Above 40 MeV, the GEF widths remain relatively close to the values obtained by carrying forward the highest available \(E_n\le20\) MeV width for each target, whereas linear extrapolation of the low-energy trend performs substantially worse. The high-energy GEF width therefore evolves comparatively slowly rather than remaining strictly constant.
The behavior of the 30- and 40-MeV models is consistent with recovering this calculated high-energy response as the training domain approaches the prediction region.

Because only \(^{242}\mathrm{Pu}\) contributes calculated training spectra above 40 MeV, the full-range dataset provides limited nuclear coverage at the highest incident energies. The high-energy results therefore primarily establish the importance of incident-energy support within the present GEF-trained domain.

\subsection{Comparison with experiment and observable dependence}

Comparison with the 60 measured post-neutron TKE distributions shows no uniform ordering among GEF, MDN, and BMDN across the five nuclei. The model giving the smallest numerical error changes with both nucleus and observable, indicating that improved reproduction of the underlying GEF calculation does not necessarily translate into uniformly improved agreement with experiment.

The relative performance also depends on the observable considered. Mean-TKE and width errors do not improve together, as illustrated by the high-energy \(^{240}\mathrm{Pu}\) comparison. More generally, the mean TKE does not exhibit the same dependence on training-energy range as the distribution width. The two observables probe different aspects of the evolving TKE distribution and can therefore respond differently to changes in the contributing fissioning systems.

The experimental widths used in the primary comparison are obtained from the variances reported in the original studies, whereas the GEF and model widths are calculated from the corresponding normalized spectra. Repeating the comparison with a common second-moment definition changes the absolute width errors but preserves the large improvement obtained when the training range is extended from 20 to 40 MeV. The principal training-energy conclusion is therefore insensitive to this difference in width definition.

\subsection{Predictive uncertainty, reproducibility, and limitations}

For BMDN, the conditional-mixture contribution dominates the modeled predictive variance in both the four-model and extended-domain analyses. Replacing the trainable prior with a fixed standard-normal prior leaves the principal high-energy width result essentially unchanged, indicating that the observed training-range dependence is not specific to the chosen prior parameterization.

The pointwise predictive intervals show a similar dependence on incident-energy coverage. Their high-energy coverage relative to GEF is poor when training is restricted to \(E_n\le20\) MeV and improves substantially as the training range is extended. This behavior is consistent with the increasing difficulty of representing the full spectral response as the prediction point moves farther beyond the incident-energy region included in training.

Variation across independent training configurations provides complementary information on numerical reproducibility. The spread increases at high energy but remains smaller than the extrapolation error of models trained only through 20 MeV, indicating that the dominant deterioration arises from limited training-domain coverage rather than from variability among individual fits.

Several limitations define the scope of the present conclusions. The neural architecture, activation function, and three-component mixture representation were not varied systematically, and the reported model widths are calculated from the conditional-mean spectrum used throughout the analysis. At the highest energies, the experimental measurements correspond to finite incident-neutron-energy groups, whereas GEF and the machine-learning models are evaluated at the reported representative group centres. The \(E_n>40\) MeV experimental subset contains 11 spectra from five target nuclei, and only \(^{242}\mathrm{Pu}\) contributes calculated training spectra above 40 MeV. The matched 40-MeV control is also based on one predetermined size-matched selection rather than an ensemble of all possible matched subsets.

Finally, the models use only \((Z,A,E_n,T)\) and do not include explicit fission-chance information, fragment masses, neutron multiplicities, or scission-related variables. The present analysis therefore characterizes how probabilistic models learn and extrapolate the GEF TKE response over the available nuclear and incident-energy domain; connecting the observed high-energy evolution to specific fission mechanisms would require additional physical information in the model inputs and training data.

\section{Conclusions}\label{sec:conclusions}

Probabilistic machine-learning models were developed for post-neutron fission-fragment TKE distributions calculated with GEF. An initial comparison of GPR, GPR-WK, MDN, and BMDN showed that their relative performance depends on whether pointwise reconstruction, spectrum-level observables, or predictive-interval quality is considered. The extended-domain analysis then focused on MDN and BMDN over a broader range of target nuclei and incident-neutron energies and included systematic comparison with 60 measured distributions for five nuclei.

The principal result concerns the high-energy TKE-distribution width. For BMDN, the width MAE above 40 MeV decreases from 5.63 MeV when training is restricted to \(E_n\le20\) MeV to 2.70--2.53 MeV for the two matched \(E_n\le30\) MeV controls and to 1.77 MeV for the matched \(E_n\le40\) MeV case. Using all available spectra through 40 MeV gives essentially the same result, 1.76 MeV. 
The matched-size controls therefore show that the dominant improvement is associated with broader incident-energy coverage rather than simply with a larger number of training spectra. The intermediate 30-MeV results further support this training-range dependence.

This conclusion is retained under target resampling, fixed-prior BMDN calculations, leave-one-out robustness checks, and a common second-moment definition. Predictive-interval coverage likewise improves as the training range is extended. The GEF-based carry-forward diagnostic indicates that the calculated high-energy width is comparatively slowly varying for the nuclei considered. The machine-learning result is therefore best interpreted as a controlled characterization of how the learned GEF response is recovered as incident-energy support is extended.

The experimental comparison is target dependent: no uniform ordering among GEF, MDN, and BMDN is obtained across the five measured nuclei. Moreover, the mean TKE does not follow the same training-range dependence as the width, emphasizing that different characteristics of the TKE distribution must be assessed separately.

Overall, the results show that probabilistic neural models can provide continuous representations of calculated fission TKE distributions together with pointwise predictive intervals whose calibration relative to GEF improves as the training domain approaches the prediction region. For the high-energy predictions considered here, adequate incident-energy coverage of the underlying calculations is important. Extending both the experimental database and the calculated nuclear and energy coverage will be necessary to determine how far these conclusions generalize beyond the present GEF-trained domain.

\section*{Author contributions}

V.P. conceived and supervised the study; designed the extended comparison with experiment, the study of the training energy range, and the robustness analyses; coordinated the integration of the calculated and experimental data; performed the final comparative and statistical analyses; interpreted the results; and wrote and revised the manuscript. M.P. contributed to the GPR methodology, initial GPR calculations, and related data analysis. C.B. contributed to the MDN/BMDN methodology, neural-network modeling and predictions, and related data analysis. All authors reviewed and approved the manuscript.

\end{document}